\documentclass{cernyrep}
\usepackage{texnames}
\usepackage[T1]{fontenc}
\newcommand{\eqn}[1]{(\ref{#1})}
\newcommand{\be}{\begin{equation}}
\newcommand{\ee}{\end{equation}}
\newcommand{\no}{\nonumber}
\newcommand{\bel}[1]{\be\label{#1}}
\newcommand{\ba}{\begin{array}{c}}
\newcommand{\bat}{\begin{array}{cc}}
\newcommand{\bath}{\begin{array}{ccc}}
\newcommand{\ea}{\end{array}}
\newcommand{\beqn}{\begin{eqnarray}}
\newcommand{\eeqn}{\end{eqnarray}}

\newcommand{\bi}{\begin{itemize}}
\newcommand{\ei}{\end{itemize}}

\newcommand{\rms}{\rm\scriptstyle}

\newcommand{\toLow}{\stackrel{q^2 \ll M_W^2}{\,\longrightarrow\,}}

\def\gap{\;\lower3pt\hbox{$\buildrel > \over \sim$}\;}
\def\lap{\;\lower3pt\hbox{$\buildrel < \over \sim$}\;}
\newcommand{\cL}{{\cal L}}

\newcommand{\cM}{{\cal M}}
\newcommand{\cO}{{\cal O}}
\newcommand{\cP}{{\cal P}}

\newcommand{\cH}{{\cal H}}
\newcommand{\cA}{{\cal A}}
\newcommand{\cI}{{\cal I}}

\newcommand{\cJ}{{\cal J}}
\newcommand{\cC}{{\cal C}}
\newcommand{\cCP}{{\cal CP}}
\newcommand{\cT}{{\cal T}}
\newcommand{\cCPT}{{\cal CPT}}
\def\bI{\mathbf{I}}
\def\bV{\mathbf{V}}
\def\bM{\mathbf{M}}
\def\bU{\mathbf{U}}
\def\bS{\mathbf{S}}
\def\bH{\mathbf{H}}
\def\bmd{\mathbf{d}}
\def\bmu{\mathbf{u}}
\def\bml{\mathbf{l}}
\def\bmf{\mathbf{f}}

\newcommand{\e}{\mbox{\rm e}}

\def\dop{d\hskip .6pt'}
\def\up{u\hskip .6pt'}
\def\lp{\ell\hskip .2pt'}
\def\nup{\nu\hskip .7pt'}
\def\sp{s\hskip .7pt'}
\def\bp{b\hskip .6pt'}

\newcommand{\Br}{\mathrm{Br}}
\begin{document}
\title{Flavour Physics and CP Violation} 

\author{Antonio Pich}

\institute{ 
IFIC, University of Val\`encia -- CSIC,
Val\`encia, Spain}

\maketitle 

\begin{abstract}
An introductory overview of the Standard Model description of flavour
is presented. The main emphasis is put on present tests of the quark-mixing
matrix structure and the phenomenological determination of its parameters.
Special attention is given to the experimental evidences of $\cCP$ violation
and their important role in our understanding of flavour dynamics.
\end{abstract}


\section{Fermion families}
\label{sec:families}

We have learnt experimentally that there are six different quark
flavours \ $u\,$, $d\,$, $s\,$, $c\,$, $b\,$, $t\,$, three different
charged leptons \ $e\,$, $\mu\,$, $\tau$ \ and their corresponding
neutrinos \ $\nu_e\,$, $\nu_\mu\,$, $\nu_\tau\,$. We can
include all these particles into the
$SU(3)_C \otimes SU(2)_L \otimes U(1)_Y$ Standard Model (SM) framework
\cite{GL:61,WE:67,SA:69}, by organizing
them into three families of quarks and leptons:
\bel{eq:families}
\left[\bat \nu_e & u \\  e^- & \dop \ea \right]\, , \qquad\quad
\left[\bat \nu_\mu & c \\  \mu^- & \sp \ea \right]\, , \qquad\quad
\left[\bat \nu_\tau & t \\  \tau^- & \bp \ea \right]\, ,
\ee
where (each quark appears in three different colours)
\bel{eq:structure}
\left[\bat \nu^{}_i & u_i \\  \ell^-_i & \dop_i \ea \right] \quad\equiv\quad
\left(\ba \nu^{}_i \\ \ell^-_i \ea \right)_L\, , \quad
\left(\ba u_i \\ \dop_i \ea \right)_L\, , \quad \ell^-_{iR}\; , \quad
u^{\phantom{j}}_{iR}\; , \quad \dop_{iR}\; ,
\ee
plus the corresponding antiparticles. Thus, the left-handed fields
are $SU(2)_L$ doublets, while their right-handed partners transform
as $SU(2)_L$ singlets. The three fermionic families
appear to have identical properties (gauge
interactions); they differ only by their mass and their flavour
quantum number.

The fermionic couplings of the photon and the $Z$ boson are flavour
conserving, \ie the neutral gauge bosons couple to a fermion and its
corresponding antifermion. In contrast, the $W^\pm$ bosons couple
any up-type quark with all down-type quarks because the weak doublet
partner of $u_i$ turns out to be a quantum superposition of
down-type mass eigenstates:
$\dop_{i} = \sum_j \bV_{ij}\, d^{\phantom{j}}_{j}$.
This flavour mixing generates a rich variety of observable phenomena,
including $\cCP$-violation effects, which can be described in a very
successful way within the SM \cite{SM:11}.

In spite of its enormous phenomenological success, The SM
does not provide
any real understanding of flavour. We do not know yet why fermions are replicated in three (and only three) nearly identical copies. Why the pattern of masses and mixings is what it is? Are the masses the only difference among the three families? What is the origin of the SM flavour structure? Which dynamics is responsible for the observed $\cCP$ violation?
The fermionic flavour is the main source of arbitrary free parameters in the SM:
9 fermion masses, 3 mixing angles and 1 complex phase, for massless neutrinos.
7 (9) additional parameters arise with non-zero Dirac (Majorana) neutrino masses:
3 masses, 3 mixing angles and 1 (3) phases.
The problem of fermion mass generation is deeply related with the
mechanism responsible for the electroweak Spontaneous Symmetry Breaking (SSB).
Thus, the origin of these parameters lies in the
most obscure part of the SM Lagrangian: the scalar sector. Clearly, the dynamics of
flavour appears to be ``terra incognita'' which deserves a careful investigation.

The following sections contain a short overview of the quark flavour sector
and its present phenomenological status. The most relevant experimental tests are
briefly described. A more pedagogic introduction to the SM can be found in
Ref.~\cite{SM:11}.

\section{Flavour structure of the Standard Model}
\label{sec:flavour}

\begin{figure}[tb]\centering
\begin{minipage}[c]{.45\linewidth}\centering
\includegraphics[width=4cm]{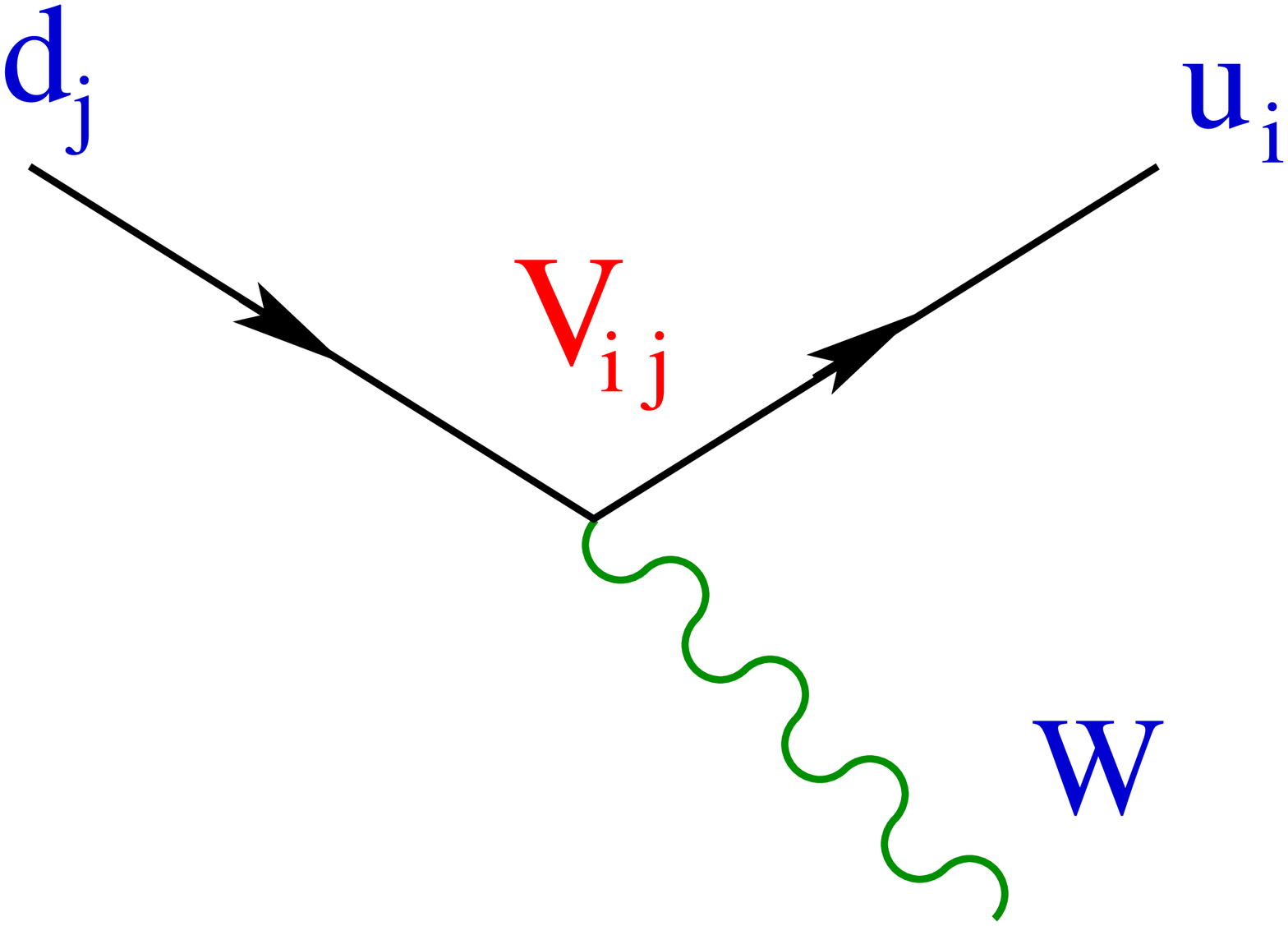}
\end{minipage}
\hskip 1cm
\begin{minipage}[c]{.45\linewidth}\centering
\includegraphics[width=4cm]{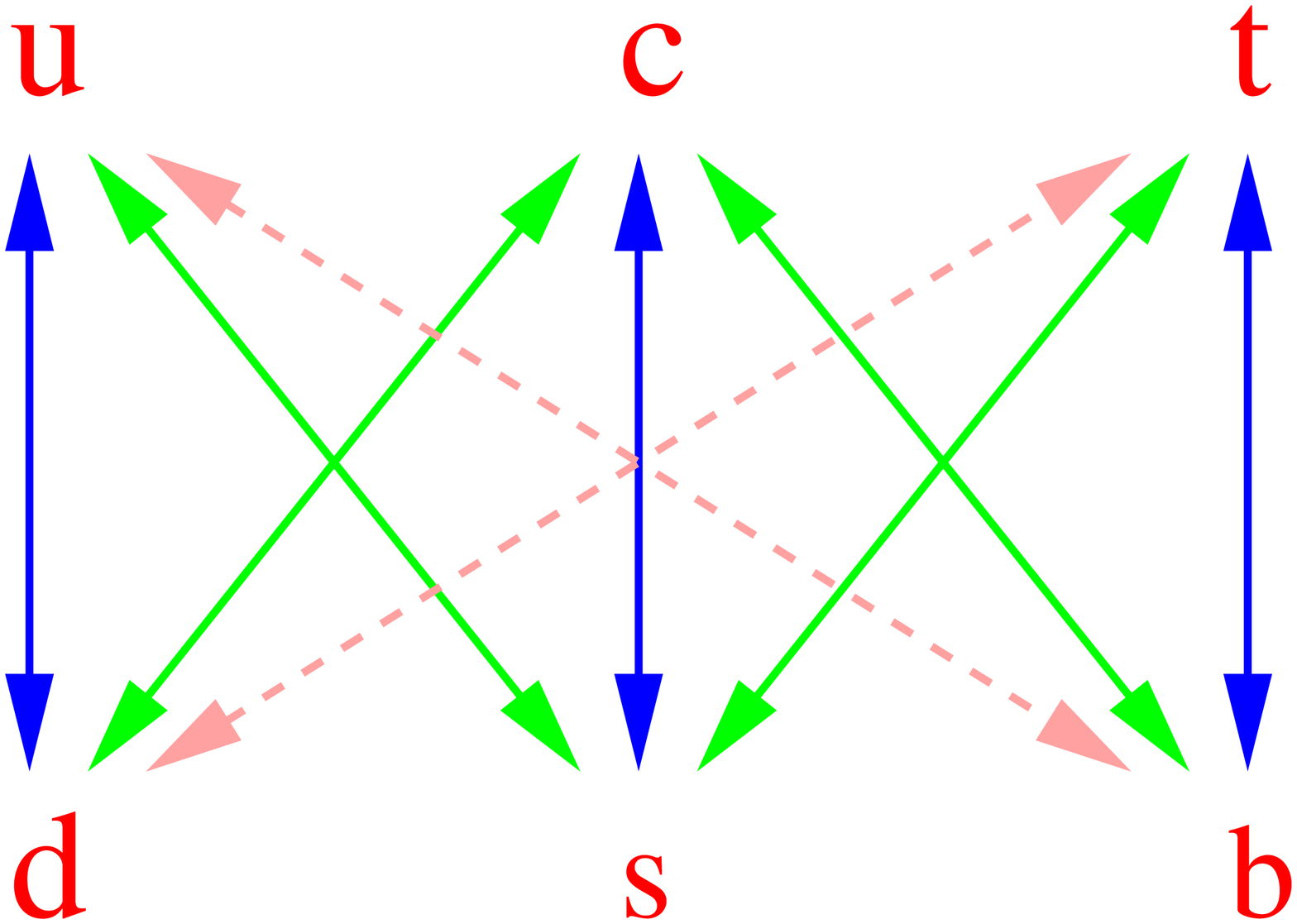}
\end{minipage}
\caption{Flavour-changing transitions through the charged-current
couplings of the $W^\pm$ bosons.}
\label{fig:CKM}
\end{figure}

In the SM flavour-changing transitions occur only in the charged-current sector
(Fig.~\ref{fig:CKM}):
\bel{eq:cc_mixing}
\cL_{\rms CC}\, = \, - {g\over 2\sqrt{2}}\,\left\{
W^\dagger_\mu\,\left[\,\sum_{ij}\;
\bar u_i\,\gamma^\mu(1-\gamma_5) \,\bV_{\! ij}\, d_j
\; +\;\sum_\ell\; \bar\nu_\ell\,\gamma^\mu(1-\gamma_5)\, \ell
\,\right]\, + \, \mathrm{h.c.}\right\}\, .
\ee
The so-called Cabibbo--Kobayashi--Maskawa (CKM) matrix $\bV$
\cite{cabibbo,KM:73} is generated by the same Yukawa couplings giving
rise to the quark masses.
Before SSB, there is no mixing among the different quarks, \ie $\bV = \bI$.
In order to understand the origin of the matrix $\bV$ ,
let us consider the general case of $N_G$ generations of fermions,
and denote $\nup_j$, $\lp_j$, $\up_j$, $\dop_j$ the members of the
weak family $j$ \ ($j=1,\ldots,N_G$), with definite transformation
properties under the gauge group. Owing to the fermion replication,
a large variety of fermion-scalar couplings are allowed by the gauge
symmetry. The most general Yukawa Lagrangian has the form
\beqn\label{eq:N_Yukawa}
\cL_Y &=&-\,\sum_{jk}\;\left\{
\left(\bar \up_j , \bar \dop_j\right)_L \left[\, c^{(d)}_{jk}\,
\left(\ba \phi^{(+)}\\ \phi^{(0)}\ea\right)\, \dop_{kR} \; +\;
c^{(u)}_{jk}\,
\left(\ba \phi^{(0)*}\\ -\phi^{(-)}\ea\right)\, \up_{kR}\,
\right]
\right.\no\\ && \qquad\quad\left. +\;\;
\left(\bar \nup_j , \bar \lp_j\right)_L\, c^{(\ell)}_{jk}\,
\left(\ba \phi^{(+)}\\ \phi^{(0)}\ea\right)\, \lp_{kR}
\,\right\}
\; +\; \mathrm{h.c.},
\eeqn
where
$\phi^T(x)\,\equiv\,\left( \phi^{(+)}, \phi^{(0)}\right)$
is the SM scalar doublet and
$c^{(d)}_{jk}$, $c^{(u)}_{jk}$ and $c^{(\ell)}_{jk}$
are arbitrary coupling constants.
The second term involves the $\cC$-conjugate scalar field
$\phi^c(x)\equiv i\,\sigma_2\,\phi^*(x)$.

In the unitary gauge 
$\phi^T(x)\,\equiv\,\frac{1}{\sqrt{2}}\,\left( 0, v+H\right)$,
where $v$ is the electroweak vacuum expectation value and
$H(x)$ the Higgs field.
The Yukawa Lagrangian can then be written as
\bel{eq:N_Yuka}
\cL_Y\, =\, - \left(1 + {H\over v}\right)\,\left\{\,
\overline{\bmd}\hskip .6pt'_L \,\bM_d'\,
\bmd\hskip .6pt'_R \; + \;
\overline{\bmu}\hskip .6pt'_L \,\bM_u'\,
\bmu\hskip .6pt'_R
\; + \;
\overline{\bml}\hskip .2pt'_L \,\bM'_\ell\,
\bml\hskip .2pt'_R \; +\;
\mathrm{h.c.}\right\} .
\ee
Here, $\bmd\hskip .6pt'$, $\bmu\hskip .6pt'$
and $\bml\hskip .2pt'$ denote vectors in
the $N_G$-dimensional flavour space,
with components  $\dop_j$, $\up_j$ and $\lp_j$, respectively,
and the corresponding mass matrices are given by
\bel{eq:M_c_relation}
(\bM'_d)^{}_{ij}\,\equiv\, c^{(d)}_{ij}\, {v\over\sqrt{2}}\, ,\qquad
(\bM'_u)^{}_{ij}\,\equiv\, c^{(u)}_{ij}\, {v\over\sqrt{2}}\, ,\qquad
(\bM'_\ell)^{}_{ij}\,\equiv\, c^{(\ell)}_{ij}\, {v\over\sqrt{2}}\, .
\ee
The diagonalization of these mass matrices determines the mass
eigenstates $d_j$, $u_j$ and $\ell_j$,
which are linear combinations of the corresponding weak eigenstates
$\dop_j$, $\up_j$ and $\lp_j$, respectively.

The matrix $\bM_d'$ can be decomposed as\footnote{
The condition $\det{\bM'_f}\not=0$ \
($f=d,u,\ell$)
guarantees that the decomposition
$\bM'_f=\bH_f\bU_f$ is unique:
$\bU_f\equiv\bH_f^{-1}\bM_f'$.
The matrices $\bS_f$
are completely determined (up to phases)
only if all diagonal elements of $\bM_f$
are different.
If there is some degeneracy, the arbitrariness of
$\bS_f$
reflects the freedom to define the physical fields.
If $\det{\bM'_f}=0$,
the matrices $\bU_f$ and
$\bS_f$ are not
uniquely determined, unless their unitarity is explicitly imposed.}
$\bM_d'=\bH^{}_d\,\bU^{}_d=\bS_d^\dagger\, \mathbf{\cM}^{}_d
\,\bS^{}_d\,\bU^{}_d$, where \ $\bH^{}_d\equiv
\sqrt{\bM_d'\bM_d'^{\dagger}}$ is an Hermitian positive-definite
matrix, while $\bU^{}_d$ is unitary. $\bH^{}_d$ can be diagonalized
by a unitary matrix $\bS^{}_d$; the resulting matrix
$\mathbf{\cM}^{}_d$ is diagonal, Hermitian and positive definite.
Similarly, one has \ $\bM_u'= \bH^{}_u\,\bU^{}_u= \bS_u^\dagger\,
\mathbf{\cM}^{}_u\, \bS^{}_u\,\bU^{}_u$ \ and \ $\bM_\ell'=
\bH^{}_\ell\,\bU^{}_\ell= \bS_\ell^\dagger\, \mathbf{\cM}^{}_\ell
\,\bS^{}_\ell\,\bU^{}_\ell$. In terms of the diagonal mass matrices
\bel{eq:Mdiagonal}
\mathbf{\cM}^{}_d =\mathrm{diag}(m_d,m_s,m_b,\ldots)\, ,\quad
\mathbf{\cM}^{}_u =\mathrm{diag}(m_u,m_c,m_t,\ldots)\, ,\quad
\mathbf{\cM}^{}_\ell=\mathrm{diag}(m_e,m_\mu,m_\tau,\ldots)\, ,
\ee
the Yukawa Lagrangian takes the simpler form
\bel{eq:N_Yuk_diag}
\cL_Y\, =\, - \left(1 + {H\over v}\right)\,\left\{\,
\overline{\bmd}\,\mathbf{\cM}^{}_d\,\bmd \; + \;
\overline{\bmu}\, \mathbf{\cM}^{}_u\,\bmu \; + \;
\overline{\bml}\,\mathbf{\cM}^{}_\ell\,\bml \,\right\}\, ,
\ee
where the mass eigenstates are defined by
\beqn\label{eq:S_matrices}
\bmd^{}_L &\!\!\equiv &\!\!
\bS^{}_d\, \bmd\hskip .6pt'_L \, ,
\qquad\qquad  \hskip 15.4pt
\bmu^{}_L \:\equiv\; \bS^{}_u \,\bmu\hskip .6pt'_L \, ,
\qquad\qquad\,\,\,\,\,\,\,\,\,
\bml^{}_L \:\equiv\; \bS^{}_\ell \,\bml\hskip .2pt'_L \, ,
\no\\
\bmd^{}_R &\!\!\equiv &\!\!
\bS^{}_d \bU^{}_d\,\bmd\hskip .6pt'_R \, , \qquad\qquad
\bmu^{}_R \:\equiv\; \bS^{}_u\bU^{}_u\, \bmu\hskip .6pt'_R \, , \qquad\qquad
\bml^{}_R \:\equiv\; \bS^{}_\ell\bU^{}_\ell \, \bml\hskip .2pt'_R \, .
\eeqn
Note, that the Higgs couplings are proportional to the
corresponding fermions masses.

Since, \ $\overline{\bmf}\hskip .7pt'_L\, \bmf\hskip .7pt'_L =
\overline{\bmf}^{}_L \,\bmf^{}_L$ \ and \
$\overline{\bmf}\hskip .7pt'_R \,\bmf\hskip .7pt'_R =
\overline{\bmf}^{}_R \,\bmf^{}_R$ \
($f=d,u,\ell$), the form of the neutral-current part of the
$SU(3)_C\otimes SU(2)_L\otimes U(1)_Y$ Lagrangian does not change when expressed
in terms of mass eigenstates. Therefore, there are no
flavour-changing neutral currents in the SM (GIM
mechanism \cite{GIM:70}). This
is a consequence of treating all equal-charge fermions
on the same footing.
However, $\overline{\bmu}\hskip .7pt'_L \,\bmd\hskip .7pt'_L =
\overline{\bmu}^{}_L \,\bS^{}_u\,\bS_d^\dagger\,\bmd^{}_L\equiv
\overline{\bmu}^{}_L \bV\,\bmd^{}_L$. In general, $\bS^{}_u\not=
\bS^{}_d\,$; thus, if one writes the weak eigenstates in terms of
mass eigenstates, a $N_G\times N_G$ unitary mixing matrix $\bV$
appears in the quark charged-current sector as indicated in
Eq.~(\ref{eq:cc_mixing}).

If neutrinos are assumed to be massless, we can always redefine the
neutrino flavours, in such a way as to eliminate the
mixing in the lepton sector: $\overline{\mbox{\boldmath
$\nu$}}\hskip .7pt'_L\, \bml\hskip .7pt'_L =
\overline{\mbox{\boldmath $\nu$}}\hskip .7pt_L'\,
\bS^\dagger_\ell\,\bml^{}_L\equiv \overline{\mbox{\boldmath
$\nu$}}^{}_L\, \bml^{}_L$. Thus, we have lepton-flavour conservation
in the minimal SM without right-handed neutrinos.
If sterile $\nu^{}_R$ fields are included in the model, one
has an additional Yukawa term in Eq.~\eqn{eq:N_Yukawa}, giving rise
to a neutrino mass matrix \ $(\bM'_\nu)_{ij}\equiv c^{(\nu)}_{ij}\,
{v/\sqrt{2}}\,$. Thus, the model can accommodate non-zero neutrino
masses and lepton-flavour violation through a lepton mixing matrix
$\bV^{}_{\! L}$ analogous to the one present in the quark sector.
Note, however, that the total lepton number \ $L\equiv L_e + L_\mu +
L_\tau$ \ is still conserved. We know experimentally that
neutrino masses are tiny and, as shown in Table~\ref{table:LFV},
there are strong bounds on lepton-flavour violating decays.
However, we do have a clear evidence of neutrino oscillation phenomena.
Moreover, since right-handed neutrinos are singlets under
$SU(3)_C \otimes SU(2)_L \otimes U(1)_Y$, the SM gauge symmetry group
allows for a right-handed Majorana neutrino mass term,
violating lepton number by two units.
Non-zero neutrino masses clearly imply interesting new phenomena \cite{SM:11}.

\begin{table}[b]
\caption{Experimental upper limits (90\% C.L.) on
lepton-flavour-violating decays \cite{MEG:11,BE:88,LFVbabar,LFVbelle,BO:88}.}
\label{table:LFV}
\centering
\begin{tabular}{ll@{\hspace{1.05cm}}ll@{\hspace{1.05cm}}ll@{\hspace{1.05cm}}ll@{\hspace{1.05cm}}ll}
\hline\hline
 \multicolumn{6}{l}{$\Br (\mu^-\to X^-)\cdot 10^{12}$}
 \\
 $e^-\gamma$ & $2.4$ & $e^-2\gamma$ & $72$ & $e^-e^-e^+$ & $1.0$
 \\
 \hline
 \multicolumn{6}{l}{$\Br (\tau^-\to X^-)\cdot 10^{8}$}
 \\
 $\mu^-\gamma$ & $4.4$ &
 $e^-\gamma$ & $3.3$ &
 $\mu^-\mu^+\mu^-$ & $2.1$ &
 $\mu^-\pi^+\pi^-$ & $3.3$ &
 $\mu^-K^+K^-$ & $6.8$
 \\
 $\mu^-\pi^0$ & $11$ &
 $e^-\pi^0$ & $8.0$ &
 $e^-\mu^+\mu^-$ & $2.7$ &
 $e^-\pi^+\pi^-$ & $4.4$ &
 $e^-K^+K^-$ & $5.4$
 \\
 $\mu^-K_S$ & $2.3$ &
 $e^-K_S$ & $2.6$ &
 $\mu^-e^+e^-$ & $1.8$  &
 $\mu^+\pi^-\pi^-$ & $3.7$ &
 $e^-K^\pm\pi^\mp$ & 5.8
 \\
 $\mu^-\rho^0$ & $2.6$ &
 $e^-\phi$ & $3.1$ &
 $e^-e^+e^-$  & $2.7$  &
 $e^+\pi^-\pi^-$ & $8.8$ &
 $e^+K^-K^-$ & $6.0$
 \\
 $\mu^-\eta$ & $6.5$ &
 $e^-\eta$ & $9.2$ &
 $\mu^-\mu^-e^+$ & $1.7$ &
 $\mu^-\omega$ & 8.9 &
 $e^+K^-\pi^-$ & $6.7$
 \\
 $\mu^-K^{*0}$ & $5.9$ &
 $e^-K^{*0}$ & $5.9$ &
 $e^-e^-\mu^+$ & $1.5$ &
 $\Lambda\pi^-$ & $7.2$ &
 $\mu^+K^-\pi^-$ & $9.4$
 \\ \hline\hline
\end{tabular}\end{table}

The fermion masses and the quark mixing matrix $\bV$ are all
determined by the Yukawa couplings in Eq.~\eqn{eq:N_Yukawa}.
However, the coefficients $c_{ij}^{(f)}$ are not known; therefore we
have a bunch of arbitrary parameters. A general $N_G\times N_G$
unitary matrix is characterized by $N_G^2$ real parameters: \ $N_G
(N_G-1)/2$ \ moduli and \ $N_G (N_G+1)/2$ \ phases. In the case of
$\,\bV$, many of these parameters are irrelevant because we can
always choose arbitrary quark phases. Under the phase redefinitions
\ $u_i\to \e^{i\phi_i}\, u_i$ \ and \ $d_j\to\e^{i\theta_j}\, d_j$,
the mixing matrix changes as \ $\bV_{\! ij}\to \bV_{\!
ij}\,\e^{i(\theta_j-\phi_i)}$; thus, $2 N_G-1$ phases are
unobservable. The number of physical free parameters in the
quark-mixing matrix then gets reduced to $(N_G-1)^2$: \
$N_G(N_G-1)/2$ moduli and $(N_G-1)(N_G-2)/2$ phases.

In the simpler case of two generations, $\bV$ is determined by a
single parameter. One then recovers the Cabibbo rotation matrix
\cite{cabibbo}
\bel{eq:cabibbo}
\bV\, = \,
\left(\bat \cos{\theta_C} &\sin{\theta_C} \\[2pt] -\sin{\theta_C}& \cos{\theta_C}\ea
\right)\, .
\ee
With $N_G=3$, the CKM matrix is described by three angles and one
phase. Different (but equivalent) representations can be found in
the literature. The Particle data Group \cite{PDG} advocates the use
of the following one as the `standard' CKM parametrization:
\bel{eq:CKM_pdg}
\bV\, = \, \left[
\begin{array}{ccc}
c_{12}\, c_{13}  & s_{12}\, c_{13} & s_{13}\, \e^{-i\delta_{13}} \\[2pt]
-s_{12}\, c_{23}-c_{12}\, s_{23}\, s_{13}\, \e^{i\delta_{13}} &
c_{12}\, c_{23}- s_{12}\, s_{23}\, s_{13}\, \e^{i\delta_{13}} &
s_{23}\, c_{13}  \\[2pt]
s_{12}\, s_{23}-c_{12}\, c_{23}\, s_{13}\, \e^{i\delta_{13}} &
-c_{12}\, s_{23}- s_{12}\, c_{23}\, s_{13}\, \e^{i\delta_{13}} &
c_{23}\, c_{13}
\ea
\right] .
\ee
Here \ $c_{ij} \equiv \cos{\theta_{ij}}$ \ and \ $s_{ij} \equiv
\sin{\theta_{ij}}\,$, with $i$ and $j$ being generation labels
($i,j=1,2,3$). The real angles $\theta_{12}$, $\theta_{23}$ and
$\theta_{13}$ can all be made to lie in the first quadrant, by an
appropriate redefinition of quark field phases; then, \ $c_{ij}\geq
0\,$, $s_{ij}\geq 0$ \ and \ $0\leq \delta_{13}\leq 2\pi\,$.
Notice that $\delta_{13}$ is the only complex phase in the SM
Lagrangian. Therefore, it is the only possible source of $\cCP$-violation
phenomena. In fact, it was for this reason that the third generation
was assumed to exist \cite{KM:73},
before the discovery of the $b$ and the $\tau$.
With two generations, the SM could not explain the observed
$\cCP$ violation in the $K$ system.

\section{Lepton decays}

\begin{figure}[tbh]\centering
\begin{minipage}[t]{.35\linewidth}\centering
\includegraphics[height=3cm]{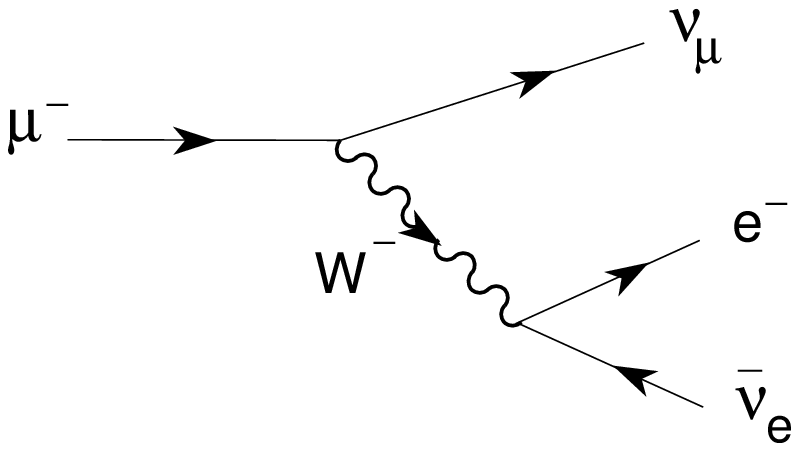}
\end{minipage}
\hskip 1.5cm
\begin{minipage}[t]{.45\linewidth}\centering
\includegraphics[height=3cm]{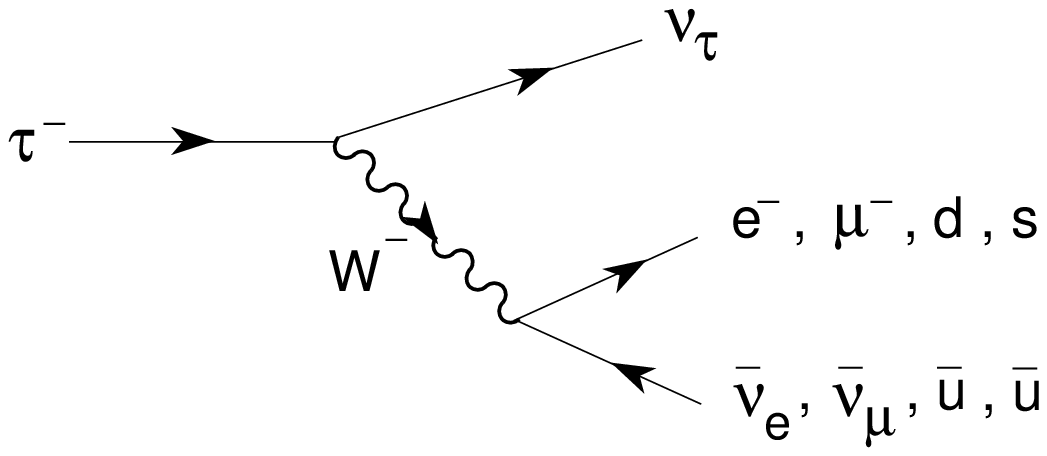}
\end{minipage}
\caption{Tree-level Feynman diagrams for \ $\mu^-\to e^-\bar\nu_e\,\nu_\mu$ \
and \ $\tau^-\to\nu_\tau X^-$ ($X^-=e^-\bar\nu_e,\, \mu^-\bar\nu_\mu,\,
d\bar u,\, s\bar u$).}
\label{fig:mu_decay}
\end{figure}

The simplest flavour-changing process is the leptonic
decay of the muon, which proceeds through the $W$-exchange
diagram shown in Fig.~\ref{fig:mu_decay}.
The momentum transfer carried by the intermediate $W$ is very small
compared to $M_W$. Therefore, the vector-boson propagator reduces
to a contact interaction,
\bel{eq:low_energy}
{-g_{\mu\nu} + q_\mu q_\nu/M_W^2 \over q^2-M_W^2}\quad\;
 \toLow\quad\; {g_{\mu\nu}\over M_W^2}\, .
\ee
The decay can then be described through an effective local
4-fermion Hamiltonian,
\bel{eq:mu_v_a}
\cH_{\mathrm{\scriptstyle eff}}\, = \,
{G_F \over\sqrt{2}}
\left[\bar e\gamma^\alpha (1-\gamma_5) \nu_e\right]\,
\left[ \bar\nu_\mu\gamma_\alpha (1-\gamma_5)\mu\right]\, ,
\ee
where
\bel{eq:G_F}
{G_F\over\sqrt{2}} \, =\, {g^2\over 8 M_W^2}\, =\,\frac{1}{2 v^2}
\ee
is called the Fermi coupling constant.
$G_F$ is fixed by the total decay width,
\bel{eq:mu_lifetime}
{1\over\tau_\mu}\, = \, \Gamma[\mu^-\to e^-\bar\nu_e\nu_\mu\, (\gamma)]
\, = \, {G_F^2 m_\mu^5\over 192 \pi^3}\,
\left( 1 + \delta_{\mathrm{\scriptstyle RC}}\right) \,
f\left(m_e^2/m_\mu^2\right) \, ,
\ee
where
$\, f(x) = 1-8x+8x^3-x^4-12x^2\ln{x}\, ,$
and
\bel{eq:qed_corr}
1+\delta_{\mathrm{\scriptstyle RC}} \, =\,
\left[1+{\alpha\over 2\pi}\left({25\over 4}-\pi^2\right)\right]\,
\left[ 1 +{3\over 5}{m_\mu^2\over M_W^2} - 2 {m_e^2\over M_W^2}\right]
\, +\,\cdots
\ee
contains the radiative higher-order corrections, which are known
to $\cO(\alpha^2)$ \cite{MS:88,vRS:99,PC:08}.
The measured lifetime \cite{MuLan:10},
$\tau_\mu=(2.196\, 980\, 3\pm 0.000\, 002\, 2)\times 10^{-6}$ s,
implies the value
\bel{eq:gf}
G_F\, = \, (1.166\, 378\, 8\pm 0.000\, 000\, 7)\times 10^{-5} \:\mbox{\rm GeV}^{-2}
\,\approx\, {1\over (293 \:\mbox{\rm GeV})^2} \, .
\ee

The decays of the $\tau$ lepton proceed through the same $W$-exchange mechanism.
The only difference is that several final states are kinematically
allowed:
$\tau^-\to\nu_\tau e^-\bar\nu_e$,
$\tau^-\to\nu_\tau\mu^-\bar\nu_\mu$,
$\tau^-\to\nu_\tau d\bar u$ and $\tau^-\to\nu_\tau s\bar u$.
Owing to the universality of the $W$ couplings, all these
decay modes have equal amplitudes (if final fermion masses and
QCD interactions are neglected), except for an additional
$N_C |\bV_{\! ui}|^2$ factor ($i=d,s$) in the semileptonic
channels, where $N_C=3$ is the number of quark colours.
Making trivial kinematical changes in Eq.~\eqn{eq:mu_lifetime},
one easily gets the lowest-order prediction for the total
$\tau$ decay width:
\bel{eq:tau_decay_width}
{1\over\tau_\tau}\equiv\Gamma(\tau) \approx
\Gamma(\mu) \left({m_\tau\over m_\mu}\right)^5
\left\{ 2 + N_C
\left( |\bV_{\! ud}|^2 + |\bV_{\! us}|^2\right)\right\}
\approx {5\over\tau_\mu}\left({m_\tau\over m_\mu}\right)^5
  ,
\ee
where we have used the CKM unitarity relation
$|\bV_{\! ud}|^2 + |\bV_{\! us}|^2 = 1 - |\bV_{\! ub}|^2
\approx 1$
(we will see later that this is an excellent approximation).
From the measured muon lifetime, one has then
$\tau_\tau\approx 3.3\times 10^{-13}$~s, to be compared
with the experimental value \cite{PDG}
$\tau_\tau^{\mathrm{exp}} = (2.906\pm 0.010)\times 10^{-13}$ s.
The numerical difference is due to the effect of QCD corrections which
enhance the hadronic $\tau$ decay width by about 20\%.
The size of these corrections has been accurately predicted
in terms of the strong coupling \cite{BNP:92}, allowing us
to extract from $\tau$ decays one of the most precise determinations of $\alpha_s$
\cite{PIalpha:11}.

In the SM all lepton doublets have identical couplings to the $W$ boson. Comparing
the measured decay widths of leptonic or semileptonic decays which only differ in the lepton flavour, one can test experimentally that
the W interaction is indeed the same, \ie that $g_e = g_\mu=g_\tau\equiv g$.
As shown in Table~\ref{tab:ccuniv}, the present data verify the universality of the leptonic charged-current couplings to the 0.2\% level.

\begin{table}[bth]\centering
\caption{Experimental determinations of the ratios \ $g_\ell/g_{\ell'}$
\cite{PDG,Babar:11,NA62:11,Flavianet:10}.}
\renewcommand{\arraystretch}{1.2}
\begin{tabular}{c@{\hspace{0.9cm}}ccccc}
 \hline\hline &
 $\Gamma_{\tau\to\mu}/\Gamma_{\tau\to e}$ &
 $\Gamma_{\pi\to\mu} /\Gamma_{\pi\to e}$ &
 $\Gamma_{K\to\mu} /\Gamma_{K\to e}$ &
 $\Gamma_{K\to\pi\mu} /\Gamma_{K\to\pi e}$ &
 $\Gamma_{W\to\mu} /\Gamma_{W\to e}$
 \\ \hline
 $|g_\mu/g_e|$
 & $1.0018\; (14)$ & $1.0021\; (16)$ & $0.998\; (2)$ & $1.001\; (2)$ & $0.991\; (9)$
 \\ \hline\hline &
 $\Gamma_{\tau\to e}/\Gamma_{\mu\to e}$ &
 $\Gamma_{\tau\to\pi}/\Gamma_{\pi\to\mu}$ &
 $\Gamma_{\tau\to K}/\Gamma_{K\to\mu}$ &
 $\Gamma_{W\to\tau}/\Gamma_{W\to\mu}$
 \\ \hline
 $|g_\tau/g_\mu|$
 & $1.0007\; (22)$ & $0.992\; (4)$ & $0.982\; (8)$ & $1.032\; (12)$
 \\ \hline\hline &
 $\Gamma_{\tau\to\mu}/\Gamma_{\mu\to e}$
 & $\Gamma_{W\to\tau}/\Gamma_{W\to e}$
 \\ \hline
 $|g_\tau/g_e|$
 & $1.0016\; (21)$ & $1.023\; (11)$
 \\ \hline\hline
\end{tabular}
\label{tab:ccuniv}
\end{table}

\section{Quark mixing}

\begin{figure}[tbh]\centering
\includegraphics[width=12cm]{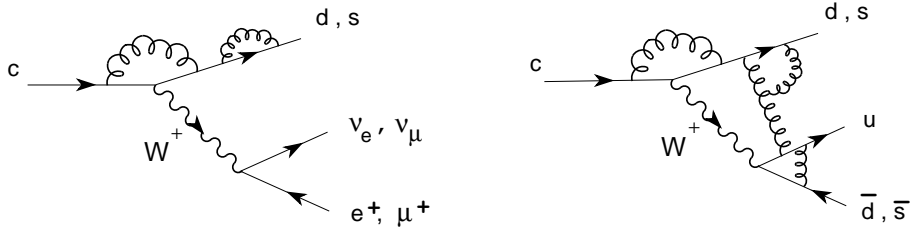}
\vskip -.3cm
\caption{
$\bV_{\! ij}$ are measured
in semileptonic decays (left), where
a single quark current is present.
Hadronic decays (right) involve two
different quark currents and are more affected
by QCD effects
(gluons can couple everywhere).
}
\label{fig:cDecay}
\end{figure}

In order to measure the CKM
matrix elements, one needs to study hadronic weak decays of the type
\ $H\to H'\, \ell^- \bar\nu_\ell$ \ or \ $H\to H'\, \ell^+ \nu_\ell$, which are
associated with the corresponding quark transitions $d_j\to u_i\,
\ell^-\bar\nu_\ell$ \  and \ $u_i\to d_j\, \ell^+\nu_\ell$
(Fig.~\ref{fig:cDecay}). Since quarks are confined within hadrons,
the decay amplitude
\bel{eq:T_decay}
T[H\to H'\, \ell^- \bar\nu_\ell]\;\approx\; {G_F\over\sqrt{2}} \;\bV_{\! ij}\;\,
\langle H'|\, \bar u_i\, \gamma^\mu (1-\gamma_5)\, d_j\, | H\rangle \;\,
\left[\,\bar \ell \,\gamma_\mu (1-\gamma_5) \,\nu_\ell\,\right]
\ee
always involves an hadronic matrix element of the weak left current.
The evaluation of this matrix element is a non-perturbative QCD
problem, which introduces unavoidable theoretical uncertainties.

One usually looks for a semileptonic transition where the matrix
element can be fixed at some kinematical point by a symmetry
principle. This has the virtue of reducing the theoretical
uncertainties to the level of symmetry-breaking corrections and
kinematical extrapolations. The standard example is a \ $0^-\to 0^-$
\ decay such as \ $K\to\pi \ell\nu_\ell\,$, $D\to K \ell\nu_\ell$ \ or \ $B\to D
\ell\nu_\ell\,$. Only the vector current can contribute in this case:
\bel{eq:vector_me}
\langle P'(k')| \,\bar u_i \,\gamma^\mu\, d_j\, | P(k)\rangle \; = \; C_{PP'}\,
\left\{\, (k+k')^\mu\, f_+(t)\, +\, (k-k')^\mu\, f_-(t)\,\right\}\, .
\ee
Here, $C_{PP'}$ is a Clebsh--Gordan factor and $t=(k-k')^2\equiv
q^2$. The unknown strong dynamics is fully contained in the form
factors $f_\pm(t)$. In the limit of equal quark masses,
$m_{u_i}=m_{d_j}$, the divergence of the vector current is zero;
thus \ $q_\mu\left[\bar u_i \gamma^\mu d_j\right] = 0$, which
implies \ $f_-(t)=0$ \ and, moreover, $f_+(0)=1$ \ to all orders in
the strong coupling because the associated flavour charge is a
conserved quantity.\footnote{
This is completely analogous to the
electromagnetic charge conservation in QED.
The conservation of the
electromagnetic current implies that the proton electromagnetic
form factor does not get any QED or QCD correction at $q^2=0$ and,
therefore,
$Q(p)=2\, Q(u)+Q(d)=|Q(e)|$. A detailed proof can be found in
Ref.~\cite{PI:96}.}
Therefore, one only needs to estimate the corrections induced by the
quark mass differences.

Since
$q^\mu\,\left[\bar \ell\gamma_\mu (1-\gamma_5)\nu_\ell\right]\sim m_\ell$, the contribution
of $f_-(t)$ is kinematically suppressed in the electron and muon modes.
The decay width can then be written as
\bel{eq:decay_width}
\Gamma(P\to P' l \nu)
\; =\; {G_F^2 M_P^5\over 192\pi^3}\; |\bV_{\! ij}|^2\; C_{PP'}^2\;
|f_+(0)|^2\; \cI\; \left(1+\delta_{\rms RC}\right)\, ,
\ee
where $\delta_{\rms RC}$ is an electroweak radiative
correction factor and $\cI$ denotes a phase-space integral,
which in the $m_\ell=0$ limit takes the form
\bel{eq:ps_integral}
\cI\;\approx\;\int_0^{(M_P-M_{P'})^2} {dt\over M_P^8}\;
\lambda^{3/2}(t,M_P^2,M_{P'}^2)\;
\left| {f_+(t)\over f_+(0)}\right|^2\, .
\ee
The usual procedure to determine $|\bV_{\! ij}|$ involves three steps:
\begin{enumerate}
\item Measure the shape of the $t$ distribution. This fixes 
$|f_+(t)/f_+(0)|$ and therefore determines $\cI$.
\item Measure the total decay width $\Gamma$. Since $G_F$ is already known
from $\mu$ decay, one gets then an
experimental value for the product $|f_+(0)\, \bV_{\! ij}|$.
\item Get a theoretical prediction for $f_+(0)$.
\end{enumerate}
It is important to realize that theoretical input is always needed.
Thus, the accuracy of the $|\bV_{\! ij}|$ determination is limited
by our ability to calculate the relevant hadronic parameters.

\subsection{Determination of $|\bV_{\! ud}|$ and $|\bV_{\! us}|$}

The conservation of the vector 
QCD currents in the
massless quark limit allows for precise determinations of the
light-quark mixings.
%
The most accurate measurement of $\bV_{\! ud}$ is done with
superallowed nuclear $\beta$ decays of the Fermi type ($0^+\to 0^+$),
where the nuclear matrix element
$\langle N'|\bar u\gamma^\mu d|N\rangle$
can be fixed by vector-current conservation.
The CKM factor is obtained through the relation \cite{MS:06,CMS:04},
\bel{eq:ft_value}
|\bV_{\! ud}|^2\, =\,
{\pi^3\ln{2}\over ft\, G_F^2 m_e^5\, (1+\delta_{\mathrm{\scriptstyle  RC}})}
\, = \,
{(2984.48\pm 0.05)\,\mbox{\rm s}\over ft\,(1+\delta_{\mathrm{\scriptstyle  RC}})}
\, ,
\ee
where $ft$ denotes the product of a phase-space statistical decay-rate factor  and the measured half-life.
In order to obtain $|\bV_{\! ud}|$, one needs to perform a careful
analysis of radiative corrections, 
including electroweak contributions, nuclear-structure corrections
and isospin-violating nuclear effects.
These nuclear-dependent corrections are quite large,
$\delta_{\mathrm{\scriptstyle RC}}\sim 3$--4\%, and have a crucial role in bringing
the results from different nuclei into good agreement.
The weighted average of the twenty most precise determinations yields \cite{HD:09}
\bel{eq:Vud}
|\bV_{\! ud}|\, =\, 0.97425\pm 0.00022 \, .
\ee

A nuclear-physics-independent determination can be obtained from neutron decay,
$n\to p\, e^-\bar\nu_e$.
The axial current also contributes in this case; therefore, one needs
to use the experimental value of the axial-current matrix element at $q^2=0$,
$\langle p\, | \,\bar u\gamma^\mu\gamma_5 d\, |\, n\rangle  =
G_A \;\bar p \gamma^\mu n$. The present world averages,
$g_A\equiv G_A/G_V = -1.2701\pm 0.0025$     
and
$\tau_n = (881.5\pm 1.5)$~s,    
imply \cite{MS:06,CMS:04,PDG}:
\bel{eq:n_decay}
|\bV_{\! ud}|\, =\,
\left\{ {(4908.7\pm 1.9)\,\mbox{\rm s}\over \tau_n\, (1+3 g_A^2)}\right\}^{1/2}
\, =\, 0.9765\pm 0.0018 \, ,
\ee
which is larger but less precise than \eqn{eq:Vud}.

The experimental determination of the neutron lifetime is controversial.
The most precise measurement, $\tau_n = (878.5\pm 0.8)$~s \cite{SE:05}, disagrees by $6\,\sigma$ from the 2010 PDG average $\tau_n = (885.7\pm 0.8)$~s \cite{PDG}, from which it was excluded. Since a more recent experiment \cite{Pichlmaier:10}
finds a mean life closer to the value of Ref.~\cite{SE:05}, both results have been included in the 2011 PDG average, enlarging the error with a scale factor of 2.7 to account for the discrepancies.
Including only the 3 most recent measurements \cite{SE:05,Pichlmaier:10,NI:05} leads to $\tau_n = (879.1\pm 1.2)$~s; this
implies $|\bV_{\! ud}| =0.9779\pm 0.0017$, which is 2.1 $\sigma$ larger than (\ref{eq:Vud}).
Small inconsistencies are also present in the $g_A$
measurements, with the most recent experiments \cite{AB:08} favouring slightly larger values of $|g_A|$.
Better measurements of $g_A$ and $\tau_n$ are needed.

The pion $\beta$ decay $\pi^+\to\pi^0 e^+\nu_e$ offers a
cleaner way to measure
$|\bV_{\! ud}|$. It is a pure vector transition, with very small theoretical
uncertainties. At $q^2=0$, the hadronic matrix element
does not receive isospin-breaking contributions of first order in
$m_d-m_u$, \ie $f_+(0)= 1 + \cO[(m_d-m_u)^2]$ \cite{AG:64}.
The small available phase space makes it possible to theoretically
control the form factor with high accuracy over the entire kinematical
domain \cite{CKNP:03}; unfortunately, it also implies a very suppressed
branching fraction.
From the present experimental value \cite{PiBeta:04},
Br$(\pi^+\to\pi^0 e^+\nu_e)=(1.040\pm 0.006)\times 10^{-8}$, one gets
$|\bV_{\! ud}| = 0.9741\pm 0.0002_{\rms th}\pm 0.0026_{\rms exp}$ \cite{QuarkFlavour}.
A tenfold improvement of the experimental accuracy would be needed to get
a determination competitive with \eqn{eq:Vud}.


The classic determination of $|\bV_{\! us}|$ takes advantage of the
theoretically well-understood $K_{\ell 3}$ decays. The most recent
high-statistics experiments have resulted in significant shifts in the
branching fractions and improved precision for all of the experimental inputs \cite{PDG}.
Supplemented with theoretical calculations of electromagnetic and
isospin corrections \cite{CGN:08,KN:08}, they allow us to extract the product
$|\bV_{\! us}\, f_+(0)| = 0.2163\pm 0.0005$ \cite{Flavianet:10},
with $f_+(0)= 1 + \cO[(m_s-m_u)^2]$ the vector form factor of the
$K^0\to\pi^-l^+\nu_l$ decay  \cite{AG:64,BS:60}.
The exact value of $f_+(0)$ has been thoroughly investigated since the first
precise estimate by Leutwyler and Roos, $f_+(0) = 0.961\pm 0.008$ \cite{LR:84}.
While analytical calculations based on chiral perturbation theory obtain
higher values \cite{JOP:04,CEEKPP:05}, as a consequence of including the large and positive ($\sim 0.01$) two-loop chiral corrections \cite{BT:03}, the lattice results
\cite{FLAG:11} tend to agree with the Leutwyler--Roos estimate.
Taking as reference value the most recent and precise lattice result
\cite{BO:10}, $f_+(0) = 0.960\pm 0.006$, one obtains \cite{Kaons}
\bel{eq:Vus}
|\bV_{\! us}|\, =\, 0.2255\pm 0.0005_{\rms exp}\pm 0.0012_{\rms th} \, .
\ee

Information on $\bV_{\! us}$ can also be obtained \cite{MA:04,Flavianet:10} from
the ratio of radiative inclusive decay rates
$\Gamma[K\to\mu\nu (\gamma)]/\Gamma[\pi\to\mu\nu (\gamma)]$.
With a careful treatment ot electromagnetic and isospin-violating corrections, one extracts $|\bV_{\! us}/\bV_{\! ud}|\, |F_K/F_\pi| = 0.2763 \pm 0.0005$ \cite{CN:11}.
Taking for the ratio of meson decay constants the lattice average
$F_K/F_\pi = 1.193 \pm 0.006$ \cite{FLAG:11}, one gets \cite{CN:11}
\bel{eq:Vus_Vud}
\frac{|\bV_{\! us}|}{|\bV_{\! ud}|} \, =\, 0.2316 \pm 0.0012\, .
\ee
With the value of $|\bV_{\! ud}|$ in Eq.~\eqn{eq:Vud}, this implies
$|\bV_{\! us}| = 0.2256 \pm 0.0012$.

Hyperon decays are also sensitive to $\bV_{\! us}$ \cite{CSW:03}.
Unfortunately, in weak baryon decays the theoretical control on $SU(3)$-breaking corrections is not as good as for the meson case.
A conservative estimate of these effects leads to the result
$|\bV_{\! us}| = 0.226 \pm 0.005$ \cite{MP:05}.

The accuracy of all previous determinations is limited by theoretical uncertainties.
The separate measurement of the inclusive $|\Delta S| = 0$
and $|\Delta S| = 1$ tau decay widths provides a very
clean observable to directly measure $|\bV_{\! us}|$ \cite{GJPPS:05,PI:11},
because $SU(3)$-breaking corrections are suppressed by two powers of the
$\tau$ mass. The present $\tau$ decay data imply
$|\bV_{\! us}| = 0.2166 \pm 0.0019_{\rms exp} \pm 0.0005_{\rms th}$ \cite{PI:11},
the error being dominated by the experimental uncertainties.
The central value has been shifted down by the inclusion of the most
recent BABAR and BELLE measurements, which find branching ratios
smaller than the previous world averages \cite{PDG}.
More precise data is needed to clarify this worrisome effect.
If the strangeness-changing $\tau$ decay width is measured with a 1\% precision,
the resulting $\bV_{\! us}$ uncertainty will get reduced to
around 0.6\%, \ie $\pm 0.0013$.

\subsection{Determination of $|\bV_{\! cb}|$ and $|\bV_{\! ub}|$}

In the limit of very heavy quark masses, QCD has additional flavour and spin
symmetries \cite{IW:89,GR:90,EH:90,GE:90} which can be used to make
rather precise determinations of $|\bV_{\! cb}|$, either from
exclusive decays
\cite{NE:91,LU:90} or
from the inclusive analysis of $b\to c\, \ell\,\bar\nu_\ell$ transitions.

When $m_b\gg \Lambda_{\mathrm{QCD}}$, all form factors
characterizing the decays $B\to D \ell \bar\nu_\ell$ and $B\to D^* \ell\bar\nu_\ell$
reduce to a single function \cite{IW:89}, which depends on the product of the
four-velocities of the two mesons
$w\equiv v_B^{\phantom{*}}\cdot v_{D^{(*)}}
= (M_B^2 + M_{D^{(*)}}^2 - q^2) / (2 M_B M_{D^{(*)}})$.
Heavy quark symmetry determines the normalization of the rate at $w = 1$, the
maximum momentum transfer to the leptons, because the corresponding vector current
is conserved in the limit of equal $B$ and $D^{(*)}$ velocities.
The $B\to  D^*$ mode has the additional advantage that
corrections to the infinite-mass limit are of second order in $1/m_b - 1/m_c$
at zero recoil ($w = 1$) \cite{LU:90}.
The exclusive determination of $|\bV_{\! cb}|$ is obtained from an extrapolation of the
measured spectrum to $w = 1$.
From $B\to D^* \ell\bar\nu_\ell$ data one gets
$|\bV_{\! cb}|\, {\cal F}(1) = (36.04\pm 0.52)\times 10^{-3}$, while
the measured $B\to D \ell \bar\nu_\ell$ distribution results in
$|\bV_{\! cb}|\, {\cal G}(1) = (42.3\pm 0.7\pm 1.3)\times 10^{-3}$
\cite{HFAG:10}. Lattice simulations are used to estimate the
deviations from unity of the two form factors at $w=1$; the most recent results
are
${\cal F}(1) = 0.908\pm 0.017$ \cite{BE:09} and
${\cal G}(1)= 1.074 \pm 0.018 \pm 0.016$ \cite{OK:05}.
A smaller value ${\cal F}(1) = 0.86\pm 0.03$ is obtained from
zero-recoil sum rules \cite{GMU:10}. Adopting the lattice estimates, one gets
\bel{eq:Vcb_Excl}
|\bV_{\! cb}|\, =\, \left\{
\bat (39.7\pm 0.9)\times 10^{-3} & \quad (B\to D^* \ell\bar\nu_\ell)\\[3pt]
(39.4\pm 1.6)\times 10^{-3} & \quad (B\to D \ell\bar\nu_\ell) \ea\right.
\; =\; (39.6\pm 0.8)\times 10^{-3} \, .
\ee

The inclusive determination uses the Operator Product Expansion
\cite{BI:93,MW:94}
to express the total $b\to c\, \ell\,\bar\nu_\ell$ rate and moments of the
differential energy and invariant-mass spectra in a double expansion in powers
of $\alpha_s$ and $1/m_b$ \cite{BBMU:03,GU:04,BBU:05,BF:06,BLMT:04}, which
includes terms of $O(1/m_b^3)$ and has recently been upgraded
to a complete $O(\alpha_s^2)$ calculation \cite{GA:11}.
The non-perturbative matrix elements of the corresponding local operators
are obtained from a global fit to experimental moments of inclusive
$B\to X_c\, \ell\,\bar\nu_\ell$ (lepton energy and hadronic invariant mass) and $B\to X_s\gamma$ (photon energy spectrum) observables. The combination of 66 different measurements results in \cite{HFAG:10}
\bel{eq:Vcb_Incl}
|\bV_{\! cb}|\, =\, (41.85\pm 0.73)\times 10^{-3} \, .
\ee

At present there is a $2.1\,\sigma$ discrepancy between the exclusive and inclusive
determinations. Following the PDG prescription \cite{PDG}, we average both values scaling the error by $\sqrt{\chi^2/\mathrm{dof}}=2.1$:
\bel{eq:Vcb}
|\bV_{\! cb}|\, =\, (40.8\pm 1.1)\times 10^{-3} \, .
\ee

A similar disagreement is observed for the analogous $|\bV_{\! ub}|$ determinations.
The presence of a light quark makes more difficult to control
the theoretical uncertainties. Exclusive $B\to\pi\ell\nu_\ell$ decays involve a
non-perturbative form factor $f_+(t)$ which is estimated through light-cone sum rules
\cite{BZ:05,DKMMO:08,DMOW:11} or lattice simulations \cite{DA:07,BA:09}.
The inclusive measurement requires the use of stringent experimental cuts to suppress
the $b\to X_c\ell\nu_\ell$ background; this induces large errors in the theoretical predictions \cite{QuarkFlavour,BLNP:05,AG:06,GGOU:07,ALFR:09,GNP:10,PA:09,LLM:10,GK:10},
which become sensitive to non-perturbative shape functions and depend
much more strongly on $m_b$. The PDG quotes the value \cite{PDG}
\bel{eq:Vub}
|\bV_{\! ub}|\, =\, \left\{
\bat (3.38\pm 0.36)\times 10^{-3} & \quad (B\to \pi \ell\bar\nu_\ell)\\[3pt]
(4.27\pm 0.38)\times 10^{-3} & \quad (B\to X_u \ell\bar\nu_\ell) \ea\right.
\; =\; (3.89\pm 0.44)\times 10^{-3} \, ,
\ee
where the average includes an error scaling factor $\sqrt{\chi^2/\mathrm{dof}}=1.62$.

\subsection{Determination of the charm and top CKM elements}

The analytic control of theoretical uncertainties is more difficult
in semileptonic charm decays, because the
symmetry arguments associated with the light and heavy quark limits
get corrected by sizeable symmetry-breaking effects.
The magnitudes of $|\bV_{\! cd}|$ and $|\bV_{\! cs}|$ can be extracted from
$D\to\pi\ell\nu_\ell$ and $D\to K\ell\nu_\ell$ decays.
Using the lattice determination of $f_+^{D\to\pi/K}(0)$ \cite{NDFLS:10,BE:09b},
the CLEO-c \cite{CLEO:09} measurements of these decays imply \cite{LPS:11}
\beqn\label{eq:Vcd_Vcs}
|\bV_{\! cd}|\, =\, 0.234\pm 0.026\, ,
\qquad\qquad\qquad
|\bV_{\! cs}|\, =\, 0.963\pm 0.026\, ,
\eeqn
where the errors are dominated by lattice uncertainties.

The most precise determination of $|\bV_{\! cd}|$ is based on
neutrino and antineutrino interactions. The difference of the ratio
of double-muon to single-muon production by neutrino and
antineutrino beams is proportional to the charm cross section off
valence $d$ quarks and, therefore, to $|\bV_{\! cd}|$
times the average semileptonic branching ratio of charm mesons.
Averaging data from several experiments, the PDG quotes \cite{PDG}
\bel{eq:Vcd_neutrino}
|\bV_{\! cd}|\, =\, 0.230\pm 0.011\, .
\ee
The analogous determination of $|\bV_{\! cs}|$
from $\nu s\to c X$ suffers from the uncertainty of the $s$-quark sea content.

The top quark has only been seen decaying into  bottom. From the ratio
of branching fractions
$\mathrm{Br}(t\to Wb)/\mathrm{Br}(t\to Wq)$, one determines \cite{D0:08,CDF:05}
\bel{eq:Vtb_rat}
\frac{|\bV_{\! tb}|}{\sqrt{\sum_q |\bV_{\! tq}|^2}}\, >\, 0.89\quad \mathrm{(95\% CL)}
\, ,
\ee
where $q = b, s, d$. A more direct determination of $|\bV_{\! tb}|$
can be obtained from the single top-quark production cross section,
measured by D0 \cite{D0:09} and CDF \cite{CDF:09}:
\bel{eq:Vtb}
|\bV_{\! tb}|\, =\, 0.88\pm 0.07 \, .
\ee

\subsection{Structure of the CKM matrix}

Using the previous determinations of CKM elements, we can check the unitarity of the
quark mixing matrix. The most precise test involves the elements of
the first row:
\bel{eq:unitarity_test}
|\bV_{\! ud}|^2 + |\bV_{\! us}|^2 + |\bV_{\! ub}|^2 \, = \, 1.0000\pm 0.0007 \, ,
\ee
where we have taken as reference values the determinations in Eqs.~(\ref{eq:Vud}),
(\ref{eq:Vus}) and (\ref{eq:Vub}). Radiative corrections play a crucial role at the quoted level of uncertainty, while the $|\bV_{\! ub}|^2$ contribution is negligible.

The ratio of the total
hadronic decay width of the $W$ to the leptonic one provides the sum
\cite{LEPEWWG,LEPEWWG_SLD:06}
\bel{eq:unitarity_test2}
\sum_{j\,  =\,  d, s, b}\; \left( |\bV_{\! uj}|^2 + |\bV_{\! cj}|^2\right)
\; = \; 2.002\pm 0.027\, . \ee
Although much less precise than Eq.~\eqn{eq:unitarity_test}, this
result test unitarity at the 1.3\% level. From
Eq.~\eqn{eq:unitarity_test2} one can also obtain a tighter
determination of $|\bV_{\! cs}|$, using the experimental knowledge
on the other CKM matrix elements, i.e., \ $|\bV_{\! ud}|^2 +
|\bV_{\! us}|^2 + |\bV_{\! ub}|^2 + |\bV_{\! cd}|^2 + |\bV_{\!
cb}|^2 = 1.0546\pm 0.0051\,$. This gives
\bel{eq:Vcs_LEP}
|\bV_{\! cs}| \, =\, 0.973 \pm 0.014\, ,
\ee
which is more accurate than the direct determination in Eq.~(\ref{eq:Vcd_Vcs}).

The measured entries of the CKM matrix show a hierarchical pattern, with the
diagonal elements being very close to one, the ones connecting the
two first generations having a size
\bel{eq:lambda} \lambda\approx |\bV_{\! us}| = 0.2255\pm 0.0013 \, ,
\ee
the mixing between the second and third families being of order
$\lambda^2$, and the mixing between the first and third quark generations
having a much smaller size of about $\lambda^3$.
It is then quite practical to use the
approximate parametrization \cite{WO:83}:

\bel{eq:wolfenstein}
\bV\; =\; \left[ \bath\displaystyle
1- {\lambda^2 \over 2} & \lambda & A\lambda^3 (\rho - i\eta)
\\[8pt]
-\lambda &\displaystyle 1 -{\lambda^2 \over 2} & A\lambda^2
\\[8pt]
A\lambda^3 (1-\rho -i\eta) & -A\lambda^ 2 &  1
\ea\right]\;\; +\;\; O\!\left(\lambda^4 \right) \, ,
\ee
where
\bel{eq:circle} A\approx {|\bV_{\! cb}|\over\lambda^2} = 0.802\pm
0.024 \, , \qquad\qquad \sqrt{\rho^2+\eta^2} \,\approx\,
\left|{\bV_{\! ub}\over \lambda \bV_{\! cb}}\right| \, =\, 0.423\pm
0.049 \, . \ee
Defining to all orders in $\lambda$ \cite{BLO:94}
$s_{12}\equiv\lambda$, $s_{23}\equiv A\lambda^2$ and $s_{13}\,
\e^{-i\delta_{13}}\equiv A\lambda^3 (\rho-i\eta)$,
Eq.~\eqn{eq:wolfenstein} just corresponds to a Taylor expansion of
Eq.~\eqn{eq:CKM_pdg} in powers of $\lambda$.

\section{Meson-antimeson mixing}
\label{sec:BB_mixing}

Additional information on the CKM parameters can be obtained from
flavour-changing neutral-current transitions, occurring at the 1-loop
level. An important example is provided by
the mixing between the $B^0$ meson and its antiparticle.
This process occurs through  the box diagrams
shown in Fig.~\ref{fig:Bmixing}, where
two $W$ bosons are exchanged between a pair of quark lines.
The mixing amplitude is proportional to
\bel{eq:mixing}
\langle\bar B_d^0 | \cH_{\Delta B=2} |B_0\rangle\,\sim\,
\sum_{ij}\, \bV_{\! id}^{\phantom{*}}\bV_{\! ib}^*
\bV_{\! jd}^{\phantom{*}}\bV_{\! jb}^*\; S(r_i,r_j)
\,\sim\, \bV_{\! td}^2\; S(r_t,r_t) \, ,
\ee
where $S(r_i,r_j)$ is a loop function \cite{IL:81}
which depends on the masses
($r_i\equiv m_i^2/M_W^2$)
of the up-type quarks running along the internal fermionic lines.
Owing to the unitarity of the CKM matrix, the mixing vanishes
for equal (up-type) quark masses (GIM mechanism \cite{GIM:70});
thus the effect
is proportional to the mass splittings between the $u$, $c$ and $t$ quarks.
Since the different CKM factors have all a similar size,
$\bV_{\! ud}^{\phantom{*}}\bV_{\! ub}^*\sim
\bV_{\! cd}^{\phantom{*}}\bV_{\! cb}^*\sim
\bV_{\! td}^{\phantom{*}}\bV_{\! tb}^*\sim A\lambda^3$,
the final amplitude is completely dominated by the top contribution.
This transition can then be used to perform
an indirect determination of $\bV_{\! td}$.

\begin{figure}[t]\centering
\includegraphics[width=9cm]{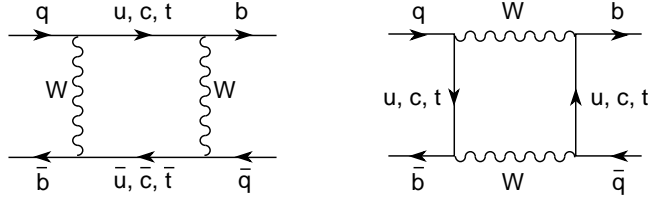}
\caption{Box diagrams contributing to $B^0$--$\bar B^0$ mixing.}
\label{fig:Bmixing}
\end{figure}

Notice that this determination has a qualitatively different character
than the ones obtained before from tree-level weak decays.
Now, we are going to test the structure of the electroweak theory at the
quantum level.
This flavour-changing transition could then be sensitive to
contributions from new physics at higher energy scales.
Moreover, the mixing amplitude crucially depends on the
unitarity of the CKM matrix.
Without the GIM mechanism embodied in the CKM mixing structure, the
calculation of the analogous $K^0\to\bar K^0$ transition (replace
the $b$ quark by a $s$ in the box diagrams) would have failed
to explain the observed $K^0$--$\bar K^0$ mixing by several orders of
magnitude \cite{GL:74}.

\subsection{General Formalism}
\label{subsec:mixing_formalism}

Since weak interactions can transform a $P^0$ state ($P=K,\, D,\, B$)
into its antiparticle $\bar P^0$, these flavour eigenstates
are not mass eigenstates and do not follow an exponential decay law.
Let us consider an arbitrary mixture of the two flavour states,
\bel{eq:mixed_state}
|\psi(t)\rangle \, =\, a(t) \, |P^0\rangle + b(t) \, |\bar P^0\rangle
\,\equiv\, \left( \ba a(t) \\ b(t) \ea\right) \, ,
\ee
with the time evolution
\bel{eq:t_eq}
i\, {d\over dt}\, |\psi(t)\rangle \, =\, \cM\, |\psi(t)\rangle \, .
\ee
Assuming $\cCPT$ symmetry to hold, the $2\times 2$ mixing matrix can be written as
\be\label{eq:mass_matrix}
{\cal M} =
\left(   \begin{array}{cc} M & M_{12} \\ M_{12}^* & M \ea   \right)
- {i\over 2}
\left(   \begin{array}{cc} \Gamma & \Gamma_{12} \\
          \Gamma_{12}^* & \Gamma \ea   \right) \, .
\ee
The diagonal elements $M$ and $\Gamma$ are real parameters,
which would correspond to the mass and width of the neutral mesons
in the absence of mixing.
The off-diagonal entries contain the {\it dispersive} and
{\it absorptive} parts of the $\Delta P=2$ transition amplitude.
If $\cCP$ were an exact symmetry, $M_{12}$ and $\Gamma_{12}$ would also be real.
The physical eigenstates of ${\cal M}$ are
\be\label{eq:eigenstates}
| P_\mp \rangle \, = \, {1\over\sqrt{|p|^2 + |q|^2}} \,
       \left[ p \, | P^0 \rangle \, \mp\, q \, | \bar P^0 \rangle \right] ,
\ee
with
\be\label{eq:q/p}
{q\over p} \, \equiv \, {1 - \bar\varepsilon \over
         1 + \bar\varepsilon} \, = \,
  \left( {M_{12}^* - {i\over 2}\Gamma_{12}^* \over
          M_{12} - {i\over 2}\Gamma_{12}} \right)^{1/2} .
\ee
If $M_{12}$ and $\Gamma_{12}$ were real then $q/p = 1$ and
$| B_\mp \rangle $ would correspond to the
$\cCP$-even and $\cCP$-odd  states
(we use the phase convention\footnote{
%
%
Since flavour is conserved by strong interactions, there is
some freedom in defining the phases of flavour eigenstates.
One could use
$\, |P^0_\zeta\rangle \equiv e^{-i\zeta} |P^0\rangle \, $ and
$|\bar P^0_\zeta\rangle \equiv e^{i\zeta} |\bar P^0\rangle$,
which satisfy
$\cC\cP\, |P^0_\zeta\rangle = - e^{-2i\zeta} |\bar P^0_\zeta\rangle$.
Both basis are trivially related:
$M_{12}^\zeta = e^{2i\zeta} M_{12}$,
$\Gamma_{12}^\zeta = e^{2i\zeta} \Gamma_{12}$ and
$(q/p)_\zeta = e^{-2i\zeta} (q/p)$.
Thus, $q/p\not=1$ does not necessarily imply $\cCP$ violation.
$\cCP$ is violated if $|q/p|\not=1$;
\ie $\mbox{\rm Re}(\bar\varepsilon)\not=0$ and
$\langle P_- |P_+\rangle \not= 0$.
Note that
$\langle P_- | P_+\rangle_\zeta =\langle P_- | P_+\rangle$.
Another phase-convention-independent quantity is
$(q/p) \, (\bar A_f/A_f)$,
where $A_f\equiv A(P^0\!\to\! f)$ and
$\bar A_f\equiv -A(\bar P^0\!\to\! f)$, for any final state $f$.}
%
%
\ $\cC\cP |P^0\rangle = - |\bar P^0\rangle$)
\bel{eq:CP_states}
|P_{1,2}\rangle\equiv {1\over\sqrt{2}} \left( |P^0\rangle\mp
|\bar P^0\rangle\right)\, , \qquad \qquad
\cC\cP\, |P_{1,2}\rangle = \pm |P_{1,2}\rangle \, .
\ee
The two mass eigenstates are no longer orthogonal when $\cCP$ is violated:
\be \langle P_- | P_+\rangle\, =\, {|p|^2-|q|^2 \over |p|^2+|q|^2}\, = \,
{2\, \mathrm{Re}\, (\bar\varepsilon)\over (1 + |\bar\varepsilon|^2)} \, .
\ee

The time evolution of a state which was originally produced
as a $P^0$ or a  $\bar P^0$  is given by
\be\label{eq:evolution}
\left( \ba | P^0(t) \rangle  \\ | \bar P^0(t) \rangle \ea \right)
\; =\;
\left( \begin{array}{cc} g_1(t)  & {q\over p}\, g_2(t) \\[5pt]
     {p\over q}\, g_2(t) & g_1(t) \ea \right)\,
\left( \ba | P^0 \rangle  \\ | \bar P^0 \rangle \ea \right) \, ,
\ee
where
\be\label{eq:g}
\left( \ba g_1(t) \\ g_2(t) \ea \right)\; =\;
\e^{-iMt}\, \e^{-\Gamma t/2}\,
\left( \ba \cos{[(\Delta M - {i\over 2} \Delta\Gamma) t/2]} \\[5pt]
   -i \sin{[(\Delta M - {i\over 2} \Delta\Gamma) t/2]} \ea \right) \, ,
\ee
with
\bel{eq:DM_DG}
\Delta M \equiv M_{P_+}-M_{P_-} \, ,\qquad\qquad
\Delta\Gamma\equiv\Gamma_{P_+}-\Gamma_{P_-}\, .
\ee

\subsection{Experimental Measurements}
\label{subsec:exp_mixing}

The main difference between the $K^0$--$\bar K^0$ and
$B^0$--$\bar B^0$ systems stems from the different kinematics involved.
The light kaon mass only allows the hadronic decay modes $K^0\to 2\pi$ and
$K^0\to 3\pi$. Since $\cC\cP\, |\pi\pi\rangle = + |\pi\pi\rangle$, the
$\cCP$-even kaon state decays into $2\pi$ whereas the
$\cCP$-odd one decays into the phase-space-suppressed $3\pi$ mode.
Therefore, there is a large lifetime difference and we have
a short-lived
$|K_S\rangle \equiv |K_-\rangle \approx
|K_1\rangle + \bar\varepsilon_K |K_2\rangle $
and a long-lived
$|K_L\rangle \equiv |K_+\rangle \approx
|K_2\rangle + \bar\varepsilon_K |K_1\rangle $
kaon,
with $\Gamma_{K_L}\ll\Gamma_{K_S}$.
One finds experimentally that
$\Delta\Gamma_{K^0}\approx -\Gamma_{K_S}\approx -2\Delta M_{K^0}$ \cite{PDG}:
\bel{eq:Kmix}
\Delta M_{K^0} = (0.5292 \pm 0.0009)\cdot 10^{10}\:\mbox{\rm s}^{-1} \, ,
\qquad\qquad
\Delta \Gamma_{K^0} = -(1.1150 \pm 0.0006)\cdot 10^{10}\:\mbox{\rm s}^{-1} \, .
\ee

In the $B$ system, there are many open decay channels and a large part of them
are common to both mass eigenstates. Therefore, the $|B_\mp\rangle $ states
have a similar lifetime; \ie $|\Delta\Gamma_{B^0}|\ll\Gamma_{B^0}$.
Moreover, whereas the $B^0$--$\bar B^0$ transition
is dominated by the top box diagram, the decay amplitudes get
obviously their main contribution from the $b\to c$ process.
Thus, $|\Delta\Gamma_{B^0} / \Delta M_{B^0}| \sim m_b^2/ m_t^2 \ll 1$.
To experimentally measure the mixing transition requires the
identification of the $B$-meson flavour at both its production and decay time.
This can be done through flavour-specific decays such as
$B^0\to X l^+\nu_l$ and $\bar B^0\to X l^-\bar\nu_l$.
In general, mixing is measured by studying pairs of $B$ mesons so that
one $B$ can be used to {\it tag} the initial flavour of the other meson.
For instance, in $e^+e^-$ machines one can look into the pair
production process
$e^+e^- \to B^0 \bar B^0 \to (X l\nu_l) \, (Y l \nu_l)$.
In the absence of mixing, the final leptons should have opposite charges;
the amount of like-sign leptons is then a clear signature of meson
mixing.

Evidence for a large $B_d^0$--$\bar B_d^0$ mixing was first
reported in 1987 by ARGUS \cite{ARGUS:87}. This provided the first indication that the
top quark was very heavy.
Since then, many experiments have analyzed the mixing probability.
The present world-average value is \cite{PDG,HFAG:10}:
\bel{eq:BdMix}
\Delta M_{B^0_d} = (0.507 \pm 0.004)\cdot 10^{12}\:\mbox{\rm s}^{-1} \, ,
\qquad\qquad
x_{B^0_d} \equiv {\Delta M_{B^0_d}\over\Gamma_{B^0_d}} = 0.771\pm 0.008
\, .
\ee
The first direct evidence of $B_s^0$--$\bar B_s^0$ oscillations was obtained by CDF
\cite{CDF:06}. The large measured mass difference reflects the CKM hierarchy
$|\bV_{\! ts}|^2 \gg |\bV_{\! td}|^2$, implying very fast oscillations
\cite{PDG,HFAG:10}:
\bel{eq:BsMix}
\Delta M_{B^0_s} = (17.77 \pm 0.12)\cdot 10^{12}\:\mbox{\rm s}^{-1} \, ,
\qquad\qquad
x_{B^0_s} \equiv {\Delta M_{B^0_s}\over\Gamma_{B^0_s}} = 26.2\pm 0.5
\, .
\ee
A more precise but still preliminary value,
$\Delta M_{B^0_s} = (17.725 \pm 0.041\pm 0.026)\:\mbox{\rm ps}^{-1}$,
has been recently obtained by LHCb  \cite{LHCb:11} using the flavour-specific decays $B^0_s\to D_s^-\pi^+$ and $\bar B^0_s\to D_s^+\pi^-$.

Evidence of mixing has been also obtained in the $D^0$--$\bar D^0$ system.
The present world averages \cite{HFAG:10},
\bel{eq:D0Mix}
x_{D^0} \equiv {\Delta M_{D^0}\over\Gamma_{D^0}} =
-\left(0.63\, {}^{+\, 0.19}_{-\, 0.20}\right)\%\, ,
\qquad\qquad
y_{D^0} \equiv {\Delta \Gamma_{D^0}\over 2\Gamma_{D^0}} =
-\left( 0.75\pm 0.12\right)\%\, ,
\ee
confirm the SM expectation of
a very slow oscillation, compared with the decay rate.

\subsection{Mixing constraints on the CKM matrix}
\label{subsec:mixing_constraints}

Long-distance contributions arising from intermediate hadronic states completely dominate the $D^0$--$\bar D^0$ mixing amplitude and are very sizeable in the $K^0$--$\bar K^0$ case, making difficult to extract useful information on the CKM matrix. The situation is much better for $B^0$ mesons, owing to the dominance of the short-distance top contribution which is known to next-to-leading order (NLO) in the strong coupling
\cite{BJW:90,HN:94}. The main uncertainty stems from the
hadronic matrix element of the $\Delta B=2$ four-quark operator
\bel{eq:DB_op}
\langle\bar B^0\, |\, (\bar b\gamma^\mu(1-\gamma_5)d)\:
(\bar b\gamma_\mu(1-\gamma_5)d)\, |\, B^0\rangle \,\equiv\,
 {8\over 3} \, M_B^2\, \xi_B^2 \, ,
\ee
which is characterized through
the non-perturbative parameter \cite{PP:95} $\xi_B(\mu)\equiv \sqrt{2}\, f_B \sqrt{B_B(\mu)}$.
Present lattice calculations obtain the ranges \cite{GDLSW:09}
$\hat\xi_{B_d} = (216\pm 15)\:\mathrm{MeV}$,
$\hat\xi_{B_s}= (266\pm 18)\:\mathrm{MeV}$ and
$\hat\xi_{B_s}/\hat\xi_{B_d}= 1.258\pm 0.033$,
where
$\hat\xi_B\approx \alpha_s(\mu)^{-3/23} \xi_B(\mu)$
is the corresponding renormalization-group-invariant quantity.
Using these values, the measured mixings in (\ref{eq:BdMix}) and (\ref{eq:BsMix}) imply
\bel{eq:V_td}
 |\bV_{\! tb}^* \bV_{\! td}|\, = \, 0.0084\pm 0.0006 \, ,
\qquad\;
 |\bV_{\! tb}^* \bV_{\! ts}|\, = \, 0.040\pm 0.003 \, ,
\qquad\;
 \frac{|\bV_{\! td}|}{|\bV_{\! ts}|}\, = \, 0.214\pm 0.006 \, .
\ee
The last number takes advantage of the smaller uncertainty in the ratio
$\hat\xi_{B_s}/\hat\xi_{B_d}$.
Since $|\bV_{\! tb}|\approx 1$, $B^0_{d,s}$ mixing provides indirect determinations of
$|\bV_{\! td}|$ and $|\bV_{\! ts}|$. The resulting value of $|\bV_{\! ts}|$ is in perfect agreement with Eq.~(\ref{eq:Vcb}), satisfying the unitarity constraint
$|\bV_{\! ts}|\approx |\bV_{\! cb}|$.
In terms of the $(\rho,\eta)$ parametrization of Eq.~\eqn{eq:wolfenstein},
one obtains
\bel{eq:circle_t}
\sqrt{(1-\rho)^2+\eta^2} \, = \, \left\{ \ba\displaystyle
\left|{\bV_{\! td}\over \lambda\bV_{\! cb}}\right|
\, = \, 0.91\pm 0.07
\\[10pt]\displaystyle
\left|{\bV_{\! td}\over \lambda\bV_{\! ts}}\right|
\, = \, 0.95\pm 0.03
\ea\right. \, .
\ee
The uncertainties in all these determinations are dominated by the theoretical errors
on the hadronic matrix elements. Therefore, they are not improved by the recent precise LHCb measurement of $\Delta M_{B^0_s}$.


\section{$\cCP$ violation}
\label{subsec:CP-Violation}

While parity and charge conjugation are violated by the weak
interactions in a maximal way, the product of the two discrete
transformations is still a good symmetry of the gauge interactions (left-handed fermions
$\leftrightarrow$ right-handed antifermions). In fact, $\cCP$
appears to be a symmetry of nearly all observed phenomena. However,
a slight violation of the $\cCP$ symmetry at the level of $0.2\%$ is
observed in the neutral kaon system and more sizeable signals of
$\cCP$ violation have been established at the B factories.
Moreover, the huge matter--antimatter asymmetry present in our
Universe is a clear manifestation of $\cCP$ violation and its
important role in the primordial baryogenesis.

The $\cCPT$ theorem guarantees that the product of the three
discrete transformations is an exact symmetry of any local and
Lorentz-invariant quantum field theory preserving micro-causality.
A violation of $\cCP$ requires then a corresponding
violation of time reversal. Since $\cT$ is an antiunitary
transformation, this requires the presence of relative complex
phases between different interfering amplitudes.

The electroweak SM Lagrangian only contains a single complex phase
$\delta_{13}$ ($\eta$). This is the sole possible source of $\cCP$
violation and, therefore, the SM predictions for $\cCP$-violating
phenomena are quite constrained. The CKM mechanism requires several
necessary conditions in order to generate an observable
$\cCP$-violation effect. With only two fermion generations, the
quark mixing mechanism cannot give rise to $\cCP$ violation;
therefore, for $\cCP$ violation to occur in a particular process,
all three generations are required to play an active role. In the
kaon system, for instance, $\cCP$ violation can only appear
at the one-loop level, where the top quark is present. In addition,
all CKM matrix elements must be non-zero and the quarks of a given
charge must be non-degenerate in mass. If any of these conditions
were not satisfied, the CKM phase could be rotated away by a
redefinition of the quark fields. $\cCP$-violation effects are then
necessarily proportional to the product of all CKM angles, and
should vanish in the limit where any two (equal-charge) quark masses
are taken to be equal. All these necessary conditions can be
summarized as a single requirement on the
original quark mass matrices $\bM'_u$ and $\bM'_d$ \cite{JA:85}:
\be
\cCP \:\mbox{\rm violation} \qquad \Longleftrightarrow \qquad
\mbox{\rm Im}\left\{\det\left[\bM_u^\prime \bM^{\prime\dagger}_u\, ,
  \,\bM^{\prime\phantom{\dagger}}_d \bM^{\prime\dagger}_d\right]\right\} \,\not=\, 0 \, .
\ee

Without performing any detailed calculation, one can make the
following general statements on the implications of the CKM mechanism
of $\cCP$ violation:

\bi
\item[--]
Owing to unitarity, for any choice of $i,j,k,l$ (between 1 and 3),
\beqn\label{eq:J_relation}
\mbox{\rm Im}\left[
\bV^{\phantom{*}}_{ij}\bV^*_{ik}\bV^{\phantom{*}}_{lk}\bV^*_{lj}\right]
\, =\, \cJ \,\sum_{m,n=1}^3 \epsilon_{ilm}\epsilon_{jkn}\, ,
\qquad\quad\\
\cJ \, =\, c_{12}\, c_{23}\, c_{13}^2\, s_{12}\, s_{23}\, s_{13}\, \sin{\delta_{13}}
\,\approx\, A^2\lambda^6\eta \, < \, 10^{-4}\, .
\eeqn
Any $\cCP$-violation observable involves the product $\cJ$
\cite{JA:85}. Thus, violations of the $\cCP$ symmetry are
necessarily small.
\item[--] In order to have sizeable $\cCP$-violating asymmetries
$\cA\equiv (\Gamma - \overline{\Gamma})/(\Gamma +
\overline{\Gamma})$, one should look for very suppressed decays,
where the decay widths already involve small CKM matrix elements.
\item[--] In the SM, $\cCP$ violation is a low-energy phenomenon,
in the sense that any effect should disappear when the quark mass
difference $m_c-m_u$ becomes negligible.
\item[--] $B$ decays are the optimal place for $\cCP$-violation signals to show up.
They involve small CKM matrix elements and are the lowest-mass
processes where the three quark generations play a direct
(tree-level) role. \ei

The SM mechanism of $\cCP$ violation is based on the unitarity of the
CKM matrix. Testing the constraints implied by unitarity
is then a way to test the source of $\cCP$ violation.
The unitarity tests in Eqs.~\eqn{eq:unitarity_test} and
\eqn{eq:unitarity_test2} involve only the moduli of the CKM parameters,
while $\cCP$ violation has to do with their phases.
More interesting are the off-diagonal unitarity conditions:
\beqn\label{eq:triangles}
\bV^\ast_{\! ud}\bV^{\phantom{*}}_{\! us} \, +\,
\bV^\ast_{\! cd}\bV^{\phantom{*}}_{\! cs} \, +\,
\bV^\ast_{\! td}\bV^{\phantom{*}}_{\! ts} & = & 0 \, ,
\nonumber\\[5pt]
\bV^\ast_{\! us}\bV^{\phantom{*}}_{\! ub} \, +\,
\bV^\ast_{\! cs}\bV^{\phantom{*}}_{\! cb} \, +\,
\bV^\ast_{\! ts}\bV^{\phantom{*}}_{\! tb} & = & 0 \, ,
\\[5pt]
\bV^\ast_{\! ub}\bV^{\phantom{*}}_{\! ud} \, +\,
\bV^\ast_{\! cb}\bV^{\phantom{*}}_{\! cd} \, +\,
\bV^\ast_{\! tb}\bV^{\phantom{*}}_{\! td} & = & 0 \, .
\nonumber\eeqn
These relations can be visualized by triangles in a complex
plane which, owing to Eq.~\eqn{eq:J_relation}, have the
same area $|\cJ|/2$.
In the absence of $\cCP$ violation, these triangles would degenerate
into segments along the real axis.

In the first two triangles, one side is much shorter than the other
two (the Cabibbo suppression factors of the three sides are
$\lambda$, $\lambda$ and $\lambda^5$ in the first triangle, and
$\lambda^4$, $\lambda^2$ and $\lambda^2$ in the second one). This is
why $\cCP$ effects are so small for $K$ mesons (first triangle), and
why certain  asymmetries in $B_s$ decays are predicted to be tiny
(second triangle).
The third triangle looks more interesting, since the three sides
have a similar size of about $\lambda^3$. They are small, which
means that the relevant $b$-decay branching ratios are small, but
once enough $B$ mesons have been produced, the $\cCP$-violation
asymmetries are sizeable. The present experimental constraints on
this triangle are shown in Fig.~\ref{fig:UTfit}, where it has been
scaled by dividing its sides by $\bV^\ast_{\!
cb}\bV^{\phantom{*}}_{\! cd}$. This aligns one side of the triangle
along the real axis and makes its length equal to 1; the coordinates
of the 3 vertices are then $(0,0)$, $(1,0)$ and
$(\bar\rho,\bar\eta)\approx (1-\lambda^2/2)\, (\rho,\eta)$.

\begin{figure}[t]\centering
\includegraphics[width=11cm,clip]{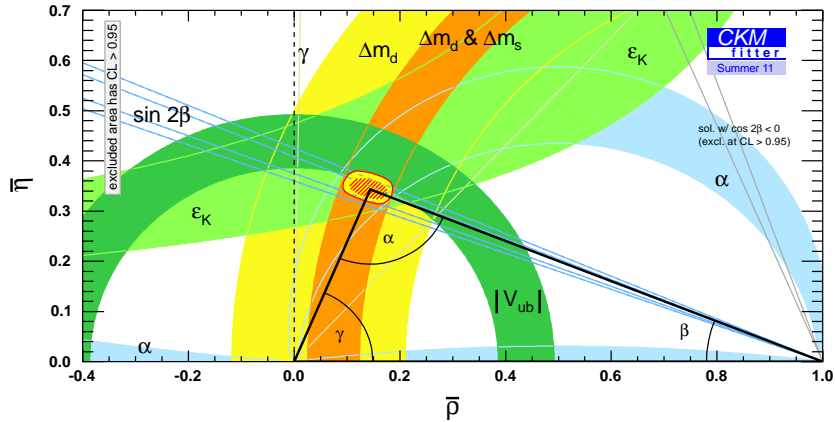}
\caption{Experimental constraints on the SM unitarity triangle
\cite{CKMfitter}.} \label{fig:UTfit}
\end{figure}

We have already determined the sides of the unitarity triangle
in Eqs.~\eqn{eq:circle} and (\ref{eq:circle_t}),
through two $\cCP$-conserving observables: $|\bV_{\! ub}/\bV_{\! cb}|$ and
$B^0_{d,s}$ mixing. This gives the circular regions shown in Fig.~\ref{fig:UTfit}, centered at the vertices $(0,0)$ and $(1,0)$. Their overlap at $\eta\not= 0$
establishes that $\cCP$ is violated (assuming unitarity).
More direct constraints on the parameter $\eta$ can be obtained from $\cCP$-violating observables, which provide sensitivity to the angles of the unitarity triangle
($\alpha + \beta + \gamma = \pi$):
\be\label{eq:angles}
\alpha\equiv\arg{\left[
   -{\bV^{\phantom{*}}_{\! td}\bV^*_{\! tb}\over
   \bV^{\phantom{*}}_{\! ud}\bV^*_{\! ub}} \right]} \, , \qquad\quad
\beta\equiv\arg{\left[
   -{\bV^{\phantom{*}}_{\! cd}\bV^*_{\! cb}\over
   \bV^{\phantom{*}}_{\! td}\bV^*_{\! tb}} \right]} \, , \qquad\quad
\gamma\equiv\arg{\left[
   -{\bV^{\phantom{*}}_{\! ud}\bV^*_{\! ub}\over
   \bV^{\phantom{*}}_{\! cd}\bV^*_{\! cb}} \right]}\, .
\ee

\subsection{Indirect and direct $\cCP$ violation in the kaon system}
\label{subsec:CP_kaon}

Any observable $\cCP$-violation effect is generated by the interference between
different amplitudes contributing to the same physical transition.
This interference can occur either through meson-antimeson mixing
or via final-state interactions, or by a combination of both effects.

The flavour-specific decays
$K^0\to\pi^- l^+\nu_l$ and $\bar K^0\to\pi^+ l^-\bar\nu_l$
provide a way to measure
the departure of the $K^0$--$\bar K^0$ mixing parameter
$|p/q|$ from unity.
In the SM, the decay amplitudes satisfy
$|A(\bar K^0\to\pi^+ l^-\bar\nu_l)| = |A(K^0\to\pi^- l^+\nu_l)|$;
therefore,
\bel{eq:deltaL}
\delta_L \,\equiv\,
{\Gamma(K_L\to\pi^- l^+\nu_l) - \Gamma(K_L\to\pi^+ l^-\bar\nu_l)\over
\Gamma(K_L\to\pi^- l^+\nu_l) + \Gamma(K_L\to\pi^+ l^-\bar\nu_l)}
\, =\, {|p|^2-|q|^2 \over |p|^2+|q|^2}
\,=\, {2\, \mathrm{Re}\, (\bar\varepsilon^{\phantom{'}}_K)\over
(1 + |\bar\varepsilon^{\phantom{'}}_K|^2)} \, .
\ee
The experimental measurement \cite{PDG},
$\delta_L = (3.32\pm 0.06)\times 10^{-3}$,
implies
\be\label{eq:Repsilon}
\mathrm{Re}\, (\bar\varepsilon^{\phantom{'}}_K)\, =\,
(1.66\pm 0.03)\times 10^{-3}\, ,
\ee
which establishes the presence of {\it indirect}\/ $\cCP$ violation
generated by the mixing amplitude.

If the flavour of the decaying meson $P$ is known,
any observed difference between the decay rate
$\Gamma(P\to f)$ and its $\cCP$ conjugate $\Gamma(\bar P\to \bar f)$
would indicate that $\cCP$ is directly violated in the decay amplitude.
One could study, for instance,
$\cCP$ asymmetries in decays such as $K^\pm\to\pi^\pm\pi^0$
where the pion charges identify the kaon flavour;
no positive signal has been found in charged kaon decays.
Since at least two interfering contributions are needed,
let us write the decay amplitudes as
\be\label{eq:direct_b}
A[P \to f] \, = \, M_1 \, e^{i\phi_1}\, e^{i \delta_1}\,
   +\, M_2 \, e^{i\phi_2}\, e^{i \delta_2} \, ,
\qquad\;
A[\bar P \to \bar f] \, = \,
  M_1\, e^{-i\phi_1} e^{i \delta_1}\, +\, M_2\, e^{-i\phi_2}e^{i \delta_2} \, ,
\ee
where $\phi_i$ denote weak phases, $\delta_i$
strong final-state phases and $M_i$ the moduli
of the matrix elements. The rate asymmetry is given by
\be\label{eq:direct_ratediff}
\mathcal{A}_{P \to f}^{\cCP}\,\equiv\,
{\Gamma[P \to f] - \Gamma[\bar P \to \bar f] \over
\Gamma[P \to f] + \Gamma[\bar P \to \bar f]} \; =\;
{-2 M_1 M_2 \,\sin{(\phi_1 - \phi_2)}\,
\sin{(\delta_1 - \delta_2)} \over
|M_1|^2 + |M_2|^2 + 2 M_1 M_2\, \cos{(\phi_1 - \phi_2)}\,
\cos{(\delta_1 - \delta_2)}} \, .
\ee
Thus, to generate a direct $\cCP$ asymmetry one needs:
1) at least two interfering amplitudes, which should be of comparable size
in order to get a sizeable asymmetry; 2) two different weak phases
[$\sin{(\phi_1 - \phi_2)}\not=0$], \ and \ 3) two different strong phases
[$\sin{(\delta_1 - \delta_2)}\not=0$].

Direct $\cCP$ violation has been searched for in decays of
neutral kaons, where $K^0$--$\bar K^0$ mixing is also involved. Thus,
both direct and indirect $\cCP$ violation need to be taken into account
simultaneously.
A $\cCP$-violation signal is provided by the ratios:
\bel{eq:etapm}
\eta_{+-} \,\equiv\, {A(K_L\to\pi^+\pi^-)\over A(K_S\to\pi^+\pi^-)}
\, =\, \varepsilon_K^{\phantom{'}} + \varepsilon_K'\, ,
\qquad\qquad
\eta_{00} \,\equiv\, {A(K_L\to\pi^0\pi^0)\over A(K_S\to\pi^0\pi^0)}
\, =\, \varepsilon_K^{\phantom{'}} - 2 \varepsilon_K'\, .
\ee
The dominant effect from $\cCP$ violation in $K^0$--$\bar{K}^0$ mixing
is contained in $\varepsilon_K^{\phantom{'}}$, while $\varepsilon_K'$ accounts for direct
$\cCP$ violation in the decay amplitudes:
\bel{eq:eps_def}
\varepsilon_K^{\phantom{'}} = \bar\varepsilon_K^{\phantom{'}} + i \xi_0 \, ,
\qquad
\varepsilon_K' = {i\over\sqrt{2}} \,\omega\, (\xi_2 - \xi_0) \, ,
\qquad
\omega \equiv {\mathrm{Re}\, (A_2)\over\mathrm{Re}\, (A_0)}\,
   e^{i(\delta_2-\delta_0)} \, ,
\qquad
\xi_I \equiv  {\mathrm{Im}\, (A_I)\over\mathrm{Re}\, (A_I)} \, .
\ee
$A_I$ and $\delta_I$ are the decay amplitudes and strong phase shifts
of isospin $I=0,2$ (these are the only two values allowed by Bose
symmetry for the final $2\pi$ state).
Although $\varepsilon_K'$ is strongly suppressed by the small ratio
$|\omega|\approx 1/22$ \cite{Kaons}, a non-zero value has been established
through very accurate measurements, demonstrating the existence of direct
$\cCP$ violation in K decays \cite{NA48,KTeV,NA31,E731}:
\bel{eq:EpsExp}
\mathrm{Re} \left(\varepsilon_K'/\varepsilon_K^{\phantom{'}}\right) =
\frac{1}{3} \left( 1
  -\left|\frac{\eta_{_{00}}}{\eta_{_{+-}}}\right|\right) =\,
(16.8 \pm 1.4) \times  10^{-4} \, .
\ee
In the SM the necessary weak phases are generated through the gluonic and electroweak penguin diagrams shown in Fig.~\ref{fig:penguin}, involving virtual up-type quarks of the 3 generations in the loop.
These short-distance contributions are known to NLO in the strong coupling
\cite{BJL:93,ciuc:93}. However, the theoretical prediction involves a delicate balance between the two isospin amplitudes and is sensitive to long-distance and isospin-violating effects. Using
chiral perturbation theory techniques \cite{PP:00,PPS:01,CPEN:03}, one finds
$\mathrm{Re} \left(\varepsilon_K'/\varepsilon_K^{\phantom{'}}\right) =
\left(19 \, {}^{+\, 11}_{-\, 9}\right) \times  10^{-4}$ \cite{Kaons},
in agreement with (\ref{eq:EpsExp}) but with a large uncertainty.

\begin{figure}[t]\centering
\includegraphics[width=4.8cm,clip]{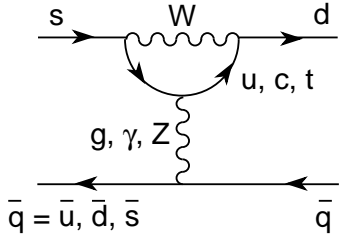}
\caption{$\Delta S=1$ penguin diagrams.}
\label{fig:penguin}
\end{figure}

Since $\mathrm{Re} \left(\varepsilon_K'/\varepsilon_K^{\phantom{'}}\right)\ll 1$, the ratios $\eta_{_{+-}}$ and $\eta_{_{00}}$
provide a measurement of
$\varepsilon_K^{\phantom{'}} = |\varepsilon_K^{\phantom{'}}|\,\e^{i\phi_\varepsilon}$ \cite{PDG}:
\bel{eq:eps_exp}
|\varepsilon_K^{\phantom{'}}| \, =\, \frac{1}{3} \left( 2 |\eta_{_{+-}}| + |\eta_{_{00}}|\right) \, =\,
(2.228\pm 0.011)\times 10^{-3} ,
\qquad\qquad
\phi_\varepsilon\, =\, (43.51\pm 0.05)^\circ\, ,
\ee
in perfect agreement with the semileptonic asymmetry $\delta_L$.
In the SM $\varepsilon_K^{\phantom{'}}$ receives short-distance contributions
from box diagrams involving virtual top and charm quarks, which are proportional
to
\bel{eq:epsilonK}
\varepsilon_K^{\phantom{'}}\,\propto\, \sum_{i,j=c,t}\,\eta_{ij}\;
\mathrm{Im}\!\left[\bV_{\! id}^{\phantom{*}}\bV_{\! is}^*
\bV_{\! jd}^{\phantom{*}}\bV_{\! js}^*\right]\, S(r_i,r_j)
\; \propto\; A^2\lambda^6 \bar\eta\,\left\{\eta_{tt}\, A^2 \lambda^4 (1-\bar\rho) + P_c\right\}
\, .
\ee
The first term shows the CKM dependence of the dominant top contribution,
$P_c$ accounts for the charm corrections \cite{BBL:96} and
the short-distance QCD corrections $\eta_{ij}$ are known to NLO \cite{BJW:90,HN:94,BG:10}.
The measured value of $|\varepsilon_K^{\phantom{'}}|$
determines an hyperbolic constraint in the $(\bar\rho,\bar\eta)$ plane, shown in Fig.~\ref{fig:UTfit}, taking into account the theoretical uncertainty in
the hadronic matrix element of the $\Delta S=2$ operator \cite{QuarkFlavour,Kaons}.


\subsection{$\cCP$ asymmetries in B decays}

The semileptonic decays
$B^0\to X^- l^+\nu_l$ and $\bar B^0\to X^+ l^-\bar\nu_l$
provide the most direct way to measure the amount of $\cCP$ violation in
the $B^0$--$\bar B^0$ mixing matrix, through
%
\beqn\label{eq:aSL}
a_{\mathrm{sl}}^q & \equiv &
{\Gamma(\bar B^0_q\to X^- l^+\nu_l) - \Gamma(B^0_q\to X^+ l^-\bar\nu_l)\over
\Gamma(\bar B^0_q\to X^- l^+\nu_l) + \Gamma(B^0_q\to X^+ l^-\bar\nu_l)}
\; =\; {|p|^4-|q|^4 \over |p|^4+|q|^4}
\;\approx\, 4\; \mathrm{Re}\, (\bar\varepsilon_{B_q^0})
\no\\[5pt]
&\approx &\frac{|\Gamma_{12}|}{|M_{12}|}\,\sin{\phi_q}
\;\approx\;\frac{|\Delta\Gamma_{B^0_q}|}{|\Delta M_{B^0_q}|}\,\tan{\phi_q}
\, .
\eeqn
This asymmetry is expected to be tiny because
$|\Gamma_{12}/M_{12}| \sim m_b^2/m_t^2 << 1$.
Moreover, there is an additional GIM suppression in the relative mixing phase
$\phi_q\equiv\arg{\left(-M_{12}/\Gamma_{12}\right)} \sim (m_c^2-m_u^2)/ m_b^2$,
implying a value of $|q/p|$  very close to 1.
Therefore, $a_{\mathrm{sl}}^q$ could be very sensitive to new $\cCP$ phases
contributing to $\phi_q$.
The present measurements give \cite{PDG,HFAG:10}
\bel{eq:aSL}
\mathrm{Re}(\bar\varepsilon_{B_d^0})\, =\, (-0.1\pm 1.4)\cdot 10^{-3}\, ,
\qquad\qquad
\mathrm{Re}(\bar\varepsilon_{B_s^0})\, =\, (-2.9\pm 1.5)\cdot 10^{-3}\, .
\ee
%
The non-zero value of
$\mathrm{Re}(\bar\varepsilon_{B_s^0})$ originates in a recent
D0 measurement \cite{D0:11} claiming the like-sign dimuon charge asymmetry
to be 3.9 $\sigma$ larger than the SM prediction
\cite{LN:97,CKMfitter:11}.
However this is not supported by the measured $\cCP$ asymmetries in
$B^0_s\to J/\psi \phi$ \cite{LHCb:11b,CDF:11b,D0:11b} and $B^0_s\to J/\psi f_0(980)$ \cite{LHCb:11b},
which are in good agreement with the SM expectations.

The large $B^0$--$\bar B^0$ mixing provides a different way to generate the
required $\cCP$-violating interference.
There are quite a few nonleptonic final states which are reachable
both from a $B^0$ and a $\bar B^0$. For these flavour non-specific decays
the $B^0$ (or $\bar B^0$) can decay directly to the given final state $f$,
or do it after the meson has been changed to its antiparticle via the
mixing process; \ie there are two different amplitudes,
$A(B^0\to f)$ and $A(B^0\to\bar B^0\to f)$, corresponding to two possible
decay paths. $\cCP$-violating effects can then result from the interference
of these two contributions.

The time-dependent decay probabilities for the decay of a neutral
$B$ meson created at the time $t_0=0$ as a pure $B^0$
($\bar B^0$) into the final state $f$ \ ($\bar f\equiv \cCP\, f$) are:
\beqn
\label{eq:decay_b}
\Gamma[B^0(t)\to f] &\!\!\propto &\!\!
{1\over 2}\, e^{-\Gamma_{B^0} t}\, \left(|A_f|^2 + |\bar A_f|^2\right)\,
\left\{ 1 + C_f\, \cos{(\Delta M_{B^0} t)} - S_f\, \sin{(\Delta M_{B^0} t)} \right\}\, ,
\nonumber\\[7pt]
\label{eq:decay_bbar}
\Gamma[\bar B^0(t)\to \bar f] &\!\!\propto &\!\!
{1\over 2}\, e^{-\Gamma_{B^0} t}\,\left( |\bar A_{\bar f}|^2 + |A_{\bar f}|^2\right)\,       \left\{ 1 - C_{\bar f}\, \cos{(\Delta M_{B^0} t)} + S_{\bar f}\, \sin{(\Delta M_{B^0} t)} \right\}\, ,
\quad\eeqn
where the tiny $\Delta\Gamma_{B^0}$ corrections have been neglected
and we have introduced the notation
$$
A_f \equiv A[B^0\to f] \, , \qquad\qquad
\bar A_f \equiv -A[\bar B^0\to f] \, , \qquad\qquad
\bar\rho_f\equiv \bar A_f / A_f \, ,
$$
\bel{eq:Bnotation}
A_{\bar f} \equiv A[B^0\to \bar f]\, , \qquad\qquad
\bar A_{\bar f} \equiv -A[\bar B^0\to \bar f] \, , \qquad\qquad
\rho_{\bar f}\equiv A_{\bar f} / \bar A_{\bar f}\, ,
\ee
$$
C_f \equiv \frac{1 - |\bar\rho_f|^2}{1 + |\bar\rho_f|^2} \, , \qquad\quad
S_f \equiv \frac{2 \,\mathrm{Im}\left( {q\over p} \,\bar\rho_f\right)}{1 + |\bar\rho_f|^2}
\, , \qquad\quad
C_{\bar f} \equiv - \frac{1 - |\rho_{\bar f}|^2}{1 + |\rho_{\bar f}|^2}\, , \qquad\quad
S_{\bar f} \equiv \frac{-2 \,\mathrm{Im}\left( {p\over q}\, \rho_{\bar f}\right)}{1 + |\rho_{\bar f}|^2}\, .
$$

$\cCP$ invariance demands the probabilities of $\cCP$-conjugate processes to be
identical. Thus, $\cCP$ conservation requires
$A_f = \bar A_{\bar f}$, $A_{\bar f} = \bar A_f$,
$\bar\rho_f = \rho_{\bar f}$ and
$\mathrm{Im}({q\over p}\, \bar\rho_f) = \mathrm{Im}({p\over q}\, \rho_{\bar f})$,
\ie $C_f = - C_{\bar f}$ and $S_f = - S_{\bar f}$.
Violation of any of the first three equalities would be a signal of
direct $\cCP$ violation. The fourth equality tests $\cCP$ violation generated
by the interference of the direct decay $B^0\to f$ and the
mixing-induced decay $B^0\to\bar B^0\to f$.

For $B^0$ mesons
%
\be
\left. {q\over p}\right|_{B^0_q} \;\approx\; \sqrt{{M_{12}^*\over M_{12}}} \;\approx\;
{\bV_{\!\! tb}^* \bV_{\!\! tq}^{\phantom{*}}
\over \bV_{\!\! tb}^{\phantom{*}} \bV_{\!\! tq}^*}
\;\equiv\; e^{-2 i \varphi^M_q} \, ,
\ee
where $\varphi^M_d =  \beta +\cO(\lambda^4)$ and $\varphi^M_s = - \beta_s+\cO(\lambda^6)$.
The angle $\beta$ is defined in Eq.~(\ref{eq:angles}), while
$\beta_s\equiv \arg{\left[ -\left(\bV^{\phantom{*}}_{\! ts}\bV^*_{\! tb}\right)/
   \left(\bV^{\phantom{*}}_{\! cs}\bV^*_{\! cb}\right)\right]} = \lambda^2\eta +\cO(\lambda^4)$
is the equivalent angle in the $B^0_s$ unitarity triangle which is predicted to be tiny.
Therefore, the mixing ratio $q/p$ is given by a known weak phase.

An obvious example of final states $f$ which can be reached both from the
$B^0$ and the $\bar B^0$ are $\cCP$ eigenstates; \ie states such that
$\bar f = \zeta_f f$ \ ($\zeta_f = \pm 1$).
In this case, $A_{\bar f} = \zeta_f A_f$,
$\bar A_{\bar f} = \zeta_f \bar A_f$,
$\rho_{\bar f}= 1/ \bar\rho_f$,
$C_{\bar f}= C_f$ and $S_{\bar f}= S_f$.
A non-zero value of $C_f$ or $S_f$ signals then $\cCP$ violation.
The ratios $\bar\rho_f$ and $\rho_{\bar f}$ depend in general on the
underlying strong dynamics. However,
for $\cCP$ self-conjugate final states, all dependence on the
strong interaction disappears if only one weak amplitude contributes to
the $B^0\to f$ and $\bar B^0\to f$ transitions \cite{CS:80,BS:81}.
In this case, we can write the decay amplitude as
$A_f = M e^{i \varphi^D} e^{i \delta_s}$, with $M = M^*$ and $\varphi^D$
and $\delta_s$  weak and strong phases.
The ratios $\bar\rho_f$ and $\rho_{\bar f}$ are then given  by
\be
\rho_{\bar f} \, =\, \bar\rho_f^*\, =\, \zeta_f\, e^{2i\varphi^D} \, .
\ee
The modulus $M$ and the unwanted strong phase cancel out completely
from these two ratios; $\rho_{\bar f}$ and $\bar\rho_f$ simplify to
a single weak phase, associated with the underlying weak quark transition.
Since $|\rho_{\bar f}| = |\bar\rho_f| = 1$,
the time-dependent decay probabilities
become much simpler. In particular, $C_f=0$ and there is no
longer any dependence on \ $\cos{(\Delta M_{B^0} t)}$.
Moreover, the coefficients of the sinusoidal terms
are then fully known in terms of CKM mixing angles only:
$S_f = S_{\bar f} = -\zeta_f\sin{[2(\varphi^M_q + \varphi^D)]}
\equiv -\zeta_f\sin{(2\Phi)}$.
In this ideal case, the time-dependent $\cCP$-violating decay asymmetry
\be
{\Gamma[\bar B^0(t)\to\bar f] - \Gamma[B^0(t)\to f] \over
 \Gamma[\bar B^0(t)\to \bar f] + \Gamma[B^0(t)\to f]} \; = \; -
\zeta_f\sin{(2 \Phi)} \, \sin{(\Delta M_{B^0} t)}
\ee
provides a direct and clean measurement of  the CKM parameters
\cite{KLPS:88}.

When several decay amplitudes with different phases
contribute, $|\bar{\rho}_f|\not=1$ and the interference term will
depend both on CKM parameters and on the strong dynamics embodied
in $\bar{\rho}_f$.
The leading contributions to
$\bar b\to\bar q' q'\bar q$ are either the tree-level $W$ exchange
or penguin topologies generated by gluon ($\gamma$, $Z$) exchange.
Although of higher order in the strong (electroweak) coupling, penguin
amplitudes are logarithmically enhanced by the virtual $W$ loop and
are  potentially competitive. Table~\ref{tab:decays} contains the CKM
factors associated with the two topologies for
different $B$ decay modes into $\cCP$ eigenstates.

\begin{table}[t]
\centering
\caption{CKM factors and relevant angle $\Phi$ for some $B$ decays into
$\cCP$ eigenstates.} \label{tab:decays}
\renewcommand{\arraystretch}{1.2}
\begin{tabular}{l@{\hspace{0.5cm}}l@{\hspace{0.5cm}}l@{\hspace{0.5cm}}l@{\hspace{0.5cm}}l}
\hline\hline
Decay & Tree-level CKM & Penguin CKM & Exclusive channels & $\Phi$ \\
\hline
$\bar b \to \bar c  c \bar s$ & $A \lambda^2$ & $-A \lambda^2$ &
$ B^0_d\to J/\psi K_S ,
J/\psi K_L$ & $\beta$ \\
&&& $ B^0_s\to D_s^+ D_s^-, J/\psi \phi$ & $-\beta_s$ \\ \hline
$\bar b\to\bar s s \bar s$ &  & $-A \lambda^2$ &
$ B^0_d\to K_S\phi, K_L\phi$ &
$\beta$ \\
&&& $B^0_s\to\phi\phi$ & $-\beta_s$ \\ \hline
$\bar b\to\bar d d\bar s$ &  & $-A \lambda^2$ &
$ B^0_s\to K_S K_S, K_L K_L$ &
$-\beta_s$ \\ \hline
$\bar b\to\bar c c\bar d$ & $-A\lambda^3$ & $A\lambda^3 (1-\rho - i \eta)$ &
$ B^0_d\to D^+ D^- , J/\psi\pi^0$ & $\approx \beta$
\\ &&& $ B^0_s\to J/\psi K_S, J/\psi K_L$ & $\approx -\beta_s$ \\ \hline
$\bar b\to\bar u u\bar d$ & $A \lambda^3 (\rho + i \eta)$ &
$A\lambda^3 (1 - \rho - i \eta)$ &
$ B^0_d\to\pi^+\pi^- , \rho^0\pi^0 ,\omega\pi^0$
& $\approx \beta+\gamma$ \\
 & & & $ B^0_s\to\rho^0 K_{S,L} ,\omega K_{S,L} ,\pi^0 K_{S,L}$
  & $\not= \gamma -\beta_s$ \\
\hline
$\bar b\to\bar s s\bar d$ &  & $A\lambda^3 (1 - \rho - i \eta)$ &
$ B^0_d\to K_S K_S, K_L K_L, \phi\pi^0$ & 0
\\ &&& $ B^0_s\to K_S\phi, K_L\phi$ & $-\beta-\beta_s$
\\ \hline\hline
\end{tabular}
\end{table}

The gold-plated decay mode is $B^0_d\to J/\psi K_S$. In addition of having a clean
experimental signature, the two topologies have the same (zero) weak phase. The
$\cCP$ asymmetry provides then a clean measurement of the mixing angle
$\beta$, without strong-interaction uncertainties.
Fig.~\ref{fig:Belle_asym} shows the most recent BELLE measurement \cite{SA:11} of
time-dependent $\bar b\to c\bar c \bar s$ asymmetries for $\cCP$-odd
($B^0_d\to J/\psi K_S$, $B^0_d\to \psi' K_S$, $B^0_d\to \chi_{c1} K_S$)
and  $\cCP$-even ($B^0_d\to J/\psi K_L$)
final states. A very nice oscillation is manifest, with opposite signs for the two
different choices of $\zeta_f=\pm 1$. Including the information obtained
from other $\bar b\to c\bar c \bar s$ decays, one gets the world average \cite{HFAG:10}:
\begin{equation}\label{eq:beta}
\sin{(2\beta)} = 0.68\pm 0.02\, .    
\end{equation}
Fitting an additional $\cos{(\Delta M_{B^0} t)}$ term in the measured asymmetries results
in $C_f=0.013\pm 0.017$ \cite{HFAG:10}, confirming the expected null result.
An independent measurement of $\sin{2\beta}$ can be obtained from
$\bar b\to s\bar s \bar s$ and $\bar b\to d\bar d \bar s$ decays, which only
receive penguin contributions and, therefore, could be more sensitive to new-physics corrections in the loop diagram. These modes give \
$\sin{(2\beta)} = 0.64\pm 0.04$ \cite{HFAG:10}, in perfect agreement with (\ref{eq:beta}).

\begin{figure}[t]\centering
\begin{minipage}[t]{.4\linewidth}\centering
\includegraphics[height=5cm]{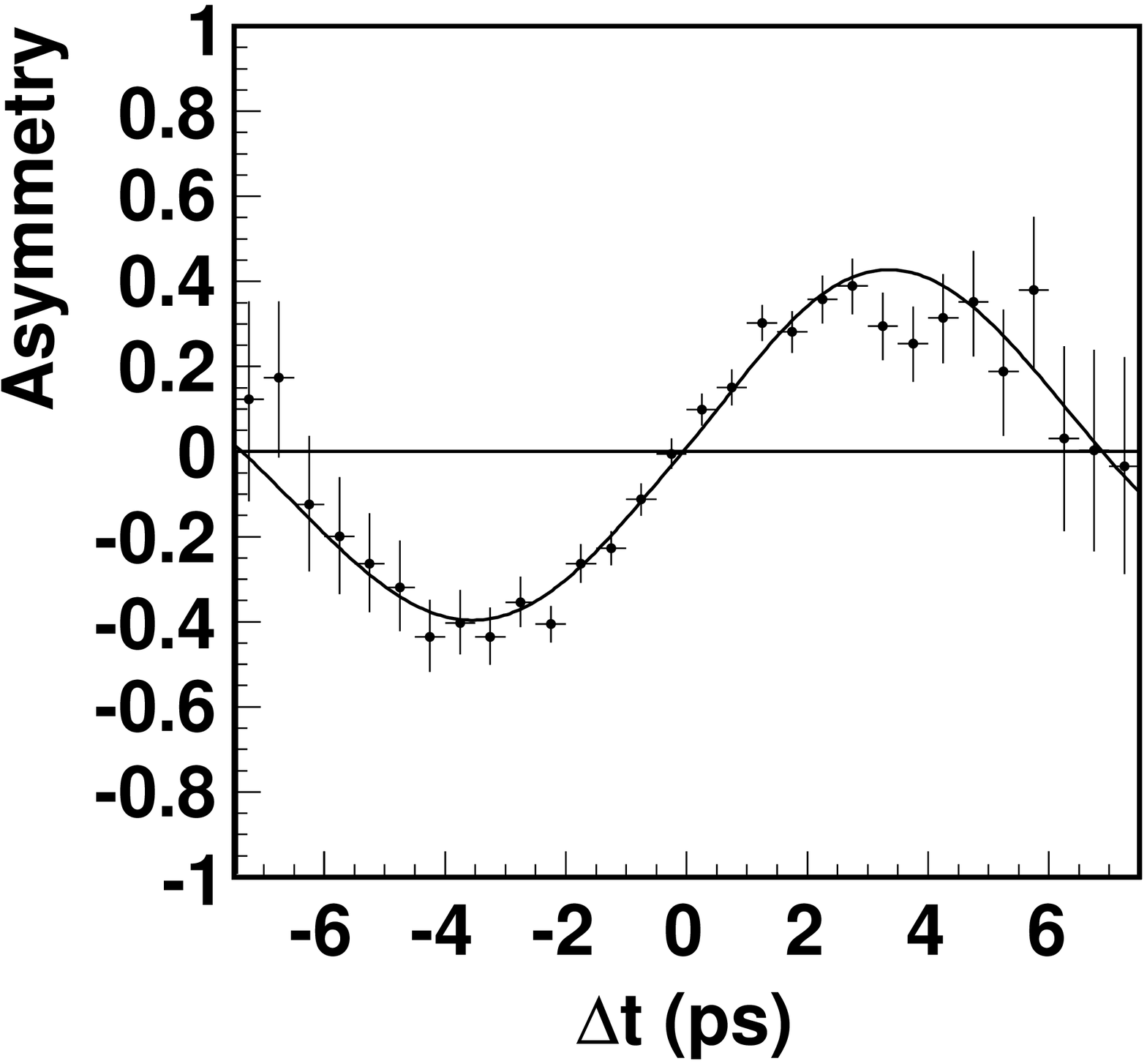}
\end{minipage}
\hskip 1.5cm
\begin{minipage}[t]{.4\linewidth}\centering
\includegraphics[height=5cm]{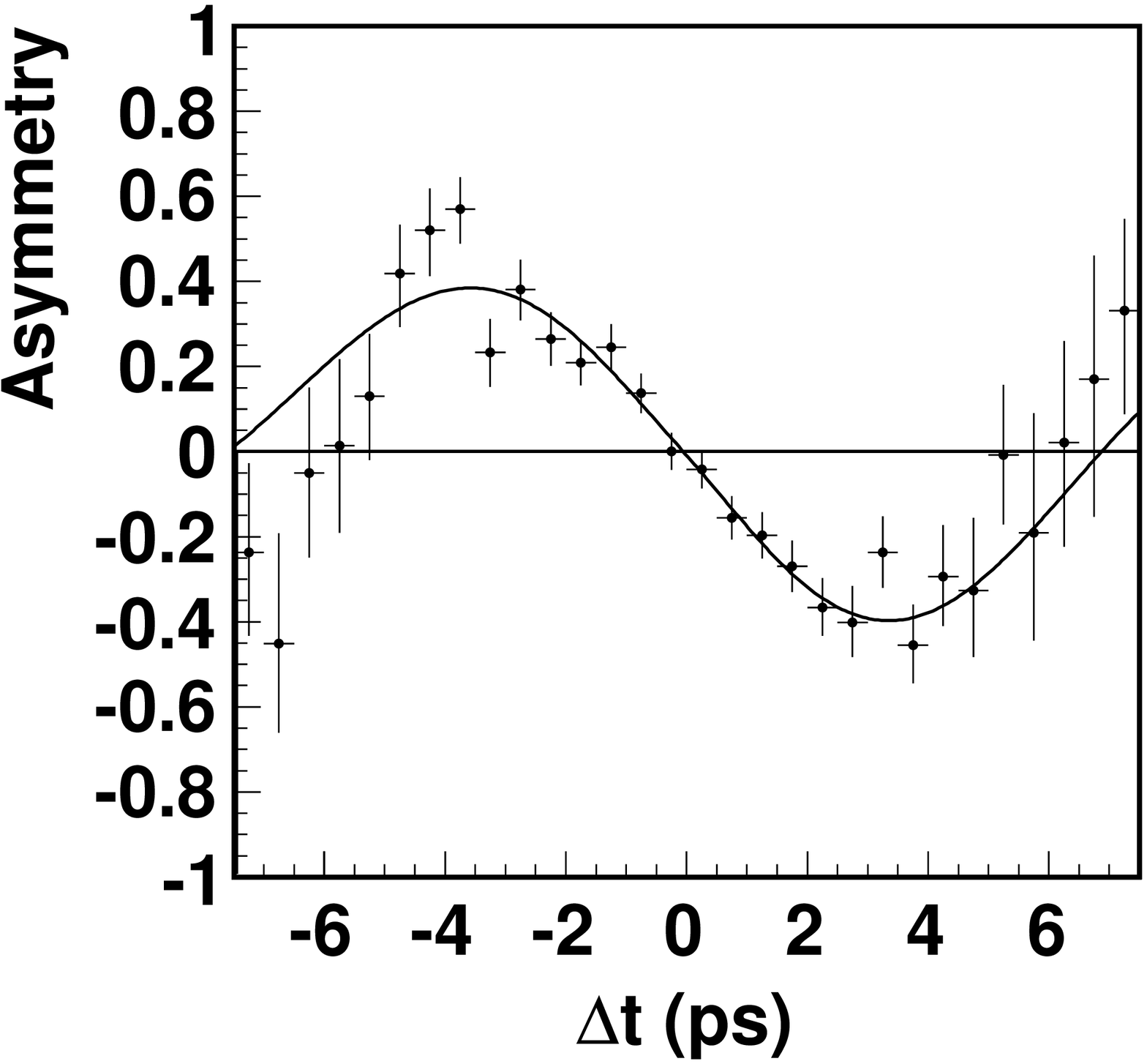}
\end{minipage}
\caption{Time-dependent asymmetries for $\cCP$-odd
($B^0_d\to J/\psi K_S$, $B^0_d\to \psi' K_S$, $B^0_d\to \chi_{c1} K_S$; $\zeta_f=-1$; left)
and  $\cCP$-even ($B^0_d\to J/\psi K_L$; $\zeta_f=+1$; right)
final states, measured by BELLE \cite{SA:11}.}
\label{fig:Belle_asym}
\end{figure}

Eq. (\ref{eq:beta}) determines the angle $\beta$ up to a four-fold ambiguity.
The time-dependent analysis of the angular $B^0_d\to J/\psi K^{*0}$ distribution and
the Dalitz plot of $B^0_d\to \bar D^0 h^0$ decays ($h^0=\pi^0,\eta,\omega$) allows us
to resolve the $\beta \leftrightarrow \frac{\pi}{2}-\beta$ ambiguity (but not the
$\beta \leftrightarrow \pi +\beta$),
showing that negative $\cos{(2\beta)}$ solutions are very unlikely \cite{Babar:05,Belle:05}.
The result fits nicely with all previous unitarity triangle constraints
in Fig.~\ref{fig:UTfit}.

A determination of $\beta+\gamma =\pi-\alpha$ can be obtained from
$\bar b\to \bar u u\bar d$ decays, such as $B^0_d\to\pi\pi$ or $B^0_d\to\rho\rho$.
However, the penguin contamination which carries a different weak phase can be
sizeable. The time-dependent asymmetry in $B^0_d\to\pi^+\pi^-$ shows indeed a
non-zero value for the $\cos{(\Delta M_{B^0} t)}$ term,
$C_f = -0.38\pm 0.06$ \cite{HFAG:10}, indicating the presence of an additional amplitude; then $S_f = -0.65\pm 0.07 \not= \sin{2\alpha}$. One could still
extract useful information on $\alpha$ (up to 16 mirror solutions),
using the isospin relations among
the $B^0_d\to\pi^+\pi^-$, $B^0_d\to\pi^0\pi^0$ and $B^+\to\pi^+\pi^0$ amplitudes and
their $\cCP$ conjugates \cite{GL:90}; however, only a loose constraint is obtained given the limited experimental precision on $B^0_d\to\pi^0\pi^0$.
Much stronger constraints are obtained from $B^0_d\to\rho\rho$ because one can use
the additional polarization information of two vectors in the final state to resolve
the different contributions and, moreover, the small branching fraction
$\mathrm{Br}(B^0_d\to\rho^0\rho^0)=(0.73\,{}^{+\, 0.27}_{-\, 0.28})\cdot 10^{-6}$
\cite{PDG} implies a very small penguin contribution.
Additional information can be obtained from $B^0_d,\bar B^0_d\to\rho^\pm\pi^\mp$,
although the final states are not $\cCP$ eigenstates.
Combining all pieces of information results in \cite{PDG,CKMfitter}
\bel{eq:alpha}
\alpha \, =\, (89.0\, {}^{+\, 4.4}_{-\, 4.2})^\circ\, .
\ee

The angle $\gamma$ cannot be determined in $\bar b\to u\bar u\bar d$ decays such as
$B_s^0\to\rho^0 K_S$ because the colour factors in the hadronic matrix element
enhance the penguin amplitude with respect to the tree-level contribution.
Instead, $\gamma$ can be measured through the tree-level decays $B^-\to D^0 K^-$
($b\to c\bar u s$) and $B^-\to \bar D^0 K^-$ ($b\to u\bar c s$), using
final states accessible in both $D^0$ and $\bar D^0$ decays and playing with
the interference of both amplitudes \cite{GL:91,GW:91,AT:97}. The sensitivity can be
optimized with Dalitz-plot analyses of $D^0,\bar D^0\to K_S\pi^+\pi^-$ decays.
The extensive studies made by BABAR and BELLE result in \cite{PDG,CKMfitter}
\bel{eq:gamma}
\gamma \, =\, (73\, {}^{+\, 22}_{-\, 25})^\circ\, .
\ee

Mixing-induced $\cCP$ violation has been also searched for
in the decays $B^0_s\to J/\psi \phi$ and $B^0_s\to J/\psi f_0(980)$. From
the corresponding time-dependent $\cCP$ asymmetries, LHCb has recently determined the angle \cite{LHCb:11b}
\bel{eq:betas}
\beta_s\; = \; -(2\pm 5)^\circ\, ,
\ee
in good agrement with the SM prediction
$\beta_s\approx \eta\lambda^2 \approx 1^\circ$.

\subsection{Global fit of the unitarity triangle}

The CKM parameters can be more accurately determined through a global fit to all available measurements, imposing the unitarity constraints and taking properly into account the theoretical uncertainties. The global fit shown in Fig.~\ref{fig:UTfit} uses frequentist statistics and gives \cite{CKMfitter}
\bel{eq:CKMfit}
\lambda\, =\, 0.2254\, {}^{+\, 0.0006}_{-\, 0.0010}\, ,
\qquad
A\, =\, 0.801\, {}^{+\, 0.026}_{-\, 0.014}\, ,
\qquad
\bar\rho\, =\, 0.144\, {}^{+\, 0.023}_{-\, 0.026}\, ,
\qquad
\bar\eta\, =\, 0.343\, {}^{+\, 0.015}_{-\, 0.014}\, .
\ee
This implies $\cJ = (2.884\, {}^{+\, 0.253}_{-\, 0.053})\cdot 10^{-5}$,
$\alpha = (90.9\, {}^{+\, 3.5}_{-\, 4.1})^\circ$,
$\beta = (21.84\, {}^{+\, 0.80}_{-\, 0.76})^\circ$ and
$\gamma = (67.3\, {}^{+\, 4.2}_{-\, 3.5})^\circ$.
Similar results are obtained by the UTfit group \cite{UTfit}, using instead a
Bayesian approach and a slightly different treatment of theoretical uncertainties.

\subsection{Direct $\cCP$ violation in $B$ and $D$ decays}

The B factories have established the presence of direct $\cCP$ violation in
several decays of $B$ mesons. The most significative signals are \cite{PDG,HFAG:10}
$$
\cA^{\cCP}_{\bar B^0_d\to K^-\pi^+} = -0.098\pm 0.013\, ,
\qquad
\cA^{\cCP}_{\bar B^0_d\to \bar K^{*0}\eta} = 0.19\pm 0.05\, ,
\qquad
\cA^{\cCP}_{\bar B^0_d\to K^{*-}\pi^+} = -0.19\pm 0.07\, ,
$$
$$
C_{B^0_d\to\pi^+\pi^-} = -0.38\pm 0.06\, ,
\qquad\qquad\qquad
\cA^{\cCP}_{B^-\to K^-D_{\cCP (+1)}} = 0.24\pm 0.06\, ,
$$
$$
\cA^{\cCP}_{B^-\to K^-\rho^0} = 0.37\pm 0.10\, ,
\qquad\qquad\qquad
\cA^{\cCP}_{B^-\to K^-\eta} = -0.37\pm 0.09\, ,
$$
\be
\cA^{\cCP}_{B^-\to K^-f_2(1270)} = -0.68\, {}^{+\, 0.19}_{-\,0.17}\, ,
\qquad\qquad\qquad
\cA^{\cCP}_{B^-\to \pi^-f_0(1370)} = 0.72\pm 0.22\, .
\ee
LHCb has also reported evidence for direct $\cCP$ violation in
$D$ decays \cite{LHCb:11c}:
\be
\cA^{\cCP}_{D^0\to K^+K^-}-\cA^{\cCP}_{D^0\to \pi^+\pi^-} = -0.82\pm 0.24\, .
\ee
Unfortunately, owing to the unavoidable presence of strong
phases, a real theoretical understanding of the corresponding SM predictions is still lacking. Progress in this direction is needed to perform meaningful tests of the CKM mechanism of $\cCP$ violation and pin down any possible effects from new physics beyond the SM framework.

\section{Rare decays}

Complementary and very valuable information could be also obtained
from rare decays which in the SM are strongly suppressed by the GIM mechanism.
These processes are sensitive to new-physics contributions with
a different flavour structure.
Well-known examples are the kaon decay modes $K^\pm\to\pi^\pm\nu\bar\nu$ and
$K_L\to\pi^0\nu\bar\nu$, where long-distance effects play a negligible role.
The decay amplitudes are dominated by short-distance loops ($Z$ penguin,
$W$ box) involving the heavy top quark, but receive also sizeable
contributions from internal charm exchanges.
The relevant hadronic matrix elements
can be obtained from $K_{l3}$ decays, assuming isospin symmetry.
The neutral decay is $\cCP$ violating and proceeds almost entirely through direct
$\cCP$ violation (via interference with mixing).
Taking the CKM matrix elements from the global fit, the predicted SM rates are
\cite{BGS:11,BSU:08,BGHN:05}:
\be
\mathrm{Br} (K^+  \to \pi^+ \nu \bar{\nu})^{\mathrm{th}}
 = (0.78 \pm 0.08) \times 10^{-10}\, ,
\qquad\;
\mathrm{Br} (K_L \to \pi^0 \nu \bar{\nu})^{\mathrm{th}}
 =  (2.4 \pm 0.4) \times 10^{-11}\, .
\ee
On the experimental side, the charged kaon mode was already observed
\cite{E949:08}, while only an upper bound on the neutral mode
has been achieved \cite{E391a:08}:
\be
\mathrm{Br} (K^+  \to \pi^+ \nu \bar{\nu})
=  (1.73^{+1.15}_{-1.05}) \times 10^{-10}\, ,
\qquad\;
\mathrm{Br} (K_L \to \pi^0 \nu \bar{\nu})
<   2.6  \times 10^{-8}  \quad  (90 \% \, {\rm C.L.})\, .
\ee
New experiments
are under development at CERN \cite{NA62:11b} and J-PARC \cite{J-PARC:10}
for charged and neutral modes, respectively, aiming
to reach $\mathcal{O}(100)$ events and begin to
seriously probe the new-physics potential of these decays.
Increased sensitivities could be obtained through the recent P996 proposal for a
$K^+  \to \pi^+ \nu \bar{\nu}$ experiment at Fermilab and the higher kaon fluxes available at Project-X \cite{TS:11}.

Another promising mode is $K_L\to\pi^0 e^+e^-$. Owing to the
electromagnetic suppression of the 2$\gamma$ $\cCP$-conserving
contribution, this decay is dominated by the $\cCP$-violating
one-photon emission amplitude. It receives contributions from
both direct and indirect $\cCP$ violation, which amount to
a predicted rate
$\mathrm{Br}(K_L \rightarrow \pi^0 e^+ e^-)^{\mathrm{th}}
= (3.1  \pm 0.9)\cdot 10^{-11}$
\cite{Kaons}, ten times smaller
than the present experimental bound
$\mathrm{Br}(K_L \rightarrow \pi^0 e^+ e^-) < 2.8 \cdot 10^{-10}$ (90\% C.L.)
\cite{KTEV:04}. Other interesting $K$ decays are discussed in Ref.~\cite{Kaons}.

The inclusive decay $\bar B\to X_s \gamma$ provides another powerful test of the SM flavour structure at the quantum loop level. The present experimental world average
 \cite{HFAG:10}
$\mathrm{Br}(\bar B\to X_s \gamma)_{E_\gamma\ge 1.6~\mathrm{GeV}} =
(3.55\pm 0.26)\cdot 10^{-4}$
agrees very well with the SM theoretical prediction \cite{MI:07}
at the next-to-next-to-leading order,
$\mathrm{Br}(\bar B\to X_s \gamma)_{E_\gamma\ge 1.6~\mathrm{GeV}}^{\mathrm{th}} =
(3.15\pm 0.23)\cdot 10^{-4}$.
Other interesting processes with $B$ mesons are $\bar B\to K^{(*)} l^+l^-$,
$\bar B^0\to l^+l^-$ and $\bar B\to K^{(*)}\nu\bar\nu$
\cite{QuarkFlavour}.

\section{Discussion}

The flavour structure of the SM is one of the main pending questions
in our understanding of weak interactions.
Although we do not know the reason of the observed family replication,
we have learnt experimentally that the number of SM generations is
just three (and no more). Therefore, we must study as precisely
as possible the few existing flavours, to get some hints on the
dynamics responsible for their observed structure.

In the SM all flavour dynamics originate in the fermion mass matrices,
which determine the measurable masses and mixings.
The SM incorporates a mechanism to generate $\cCP$ violation, through the
single phase naturally occurring in the CKM matrix.
This mechanism, deeply rooted into the unitarity structure of $\bV$,
implies very specific requirements for $\cCP$ violation to show up.
The CKM matrix has been thoroughly investigated in dedicated experiments
and a large number of $\cCP$-violating processes have been studied in detail.
At present, all flavour data seem to fit into the SM framework, confirming
that the fermion mass matrices are the dominant source of flavour-mixing phenomena.
However, a fundamental explanation of the flavour dynamics is still lacking.

The dynamics of flavour is a broad and fascinating subject, which is
closely related to the so far untested scalar sector of the SM.
LHC has already excluded a broad range of Higgs masses, narrowing down
the SM Higgs hunting to the region of low masses between 115.5 and 127 GeV (95\% CL) \cite{ATLAS,CMS}. This is precisely the range of masses preferred by precision electroweak tests \cite{SM:11}. The discovery of a neutral scalar
boson in this mass range would provide a spectacular confirmation of the SM framework.

If the Higgs boson does not show up soon, we should look for alternative mechanisms
of mass generation, satisfying the many experimental constraints which the SM has
successfully fulfilled so far. The easiest perturbative way would be enlarging the scalar sector with additional Higgs doublets, without spoiling the electroweak precision tests.
However, adding more scalar doublets to the Yukawa Lagrangian
(\ref{eq:N_Yukawa}) leads to new sources of flavour-changing phenomena; in particular,
the additional neutral scalars acquire tree-level flavour-changing neutral couplings,
which represent a major phenomenological problem. In order to satisfy the stringent experimental constraints, these scalars should be either decoupled (very large masses
above 10-50 TeV or tiny couplings) or have their Yukawa couplings aligned in flavour space \cite{PI:11b}, which suggests the existence of additional flavour symmetries at higher scales. The usual supersymmetric models with two scalar doublets avoid
this problem through a discrete symmetry which only allows one of the scalars to couple to a given right-handed fermion \cite{GW:77}; however, the presence of flavoured supersymmetric partners gives rise to other problematic sources of flavour and $\cCP$
phenomena, making necessary to tune their couplings to be tiny (or zero)
\cite{INP:10}. Models with dynamical electroweak symmetry breaking have also difficulties to accommodate the flavour constraints in a natural way.
Thus, flavour phenomena imposes severe restrictions on possible extensions of the SM.

New experimental input is expected from LHCb, BESS-III, the future Super-Belle and Super-B factories and from several kaon
(NA62, DA$\Phi$NE, $K^0$TO, TREK, KLOD, OKA, Project-X)
and muon (MEG, Mu2e, COMET, PRISM)  experiments.
These data will provide very valuable information, complementing the high-energy searches for new phenomena at LHC.
 Unexpected surprises may well be discovered, probably giving hints of new physics at higher scales and offering clues to the problems of fermion mass generation, quark
mixing and family replication.

\section*{Acknowledgements}

I want to thank the organizers for the charming atmosphere of this
school and all the students for their many interesting questions and
comments. I would also like to thank the hospitality of the Physics
Department of the Technical University of Munich, where these lectures have
been written, and the support of the Alexander von Humboldt Foundation.
This work has been supported in part by
MICINN, Spain (FPA2007-60323 and Consolider-Ingenio
2010 Program CSD2007-00042 --CPAN--) and Generalitat
Valenciana (Prometeo/2008/069).

\end{document}